\documentclass[aip,jcp,superscriptaddress,amsmath,amssymb,twocolumn,reprint]{revtex4-1}
\usepackage[version=3]{mhchem} 
\usepackage{graphicx}
\usepackage{multirow}
\usepackage{amsmath}
\usepackage{amssymb}
\usepackage{nicefrac}
\usepackage[]{natbib}
\usepackage{booktabs}
\usepackage{subcaption}
\usepackage{mathtools}
\usepackage{mwe}
\usepackage{array}
\newcolumntype{m}{>{$} c <{$}}
\usepackage{color,amscd,amsmath,amssymb,amsfonts,physics,siunitx}

\def\dd{\mathrm{d}}

\def\beq{\begin{equation}}
\def\eeq{\end{equation}}

\begin{document}     

\author{Kimberly J. Daas}
\affiliation
{Department of Chemistry \& Pharmaceutical Sciences and Amsterdam Institute of Molecular and Life Sciences (AIMMS), Faculty of Science, Vrije Universiteit, De Boelelaan 1083, 1081HV Amsterdam, The Netherlands}
\author{Derk P. Kooi}
\affiliation
{Department of Chemistry \& Pharmaceutical Sciences and Amsterdam Institute of Molecular and Life Sciences (AIMMS), Faculty of Science, Vrije Universiteit, De Boelelaan 1083, 1081HV Amsterdam, The Netherlands}
\affiliation{Microsoft Research AI4Science, Evert van de Beekstraat 354, 1118CZ Schiphol, The Netherlands}
\author{Nina C. Peters}
\affiliation
{Department of Chemistry \& Pharmaceutical Sciences and Amsterdam Institute of Molecular and Life Sciences (AIMMS), Faculty of Science, Vrije Universiteit, De Boelelaan 1083, 1081HV Amsterdam, The Netherlands}
\author{Eduardo Fabiano}
\affiliation{Institute for Microelectronics and Microsystems (CNR-IMM), Via Monteroni, Campus Unisalento, 73100 Lecce, Italy}
\author{Fabio Della Sala}
\affiliation{Institute for Microelectronics and Microsystems (CNR-IMM), Via Monteroni, Campus Unisalento, 73100 Lecce, Italy}
\author{Paola Gori-Giorgi}
\affiliation{Department of Chemistry \& Pharmaceutical Sciences and Amsterdam Institute of Molecular and Life Sciences (AIMMS), Faculty of Science, Vrije Universiteit, De Boelelaan 1083, 1081HV Amsterdam, The Netherlands}
\affiliation{Microsoft Research AI4Science,  Evert van de Beekstraat 354, 1118CZ Schiphol, The Netherlands}
\author{Stefan Vuckovic}
\email{stefan.vuckovic@unifr.ch}
\affiliation{Department of Chemistry, Faculty of Science and Medicine, Université de Fribourg/Universität Freiburg, Chemin du Musée 9, CH-1700 Fribourg, Switzerland}
\title{Regularized and Opposite spin-scaled functionals from M{\o}ller-Plesset adiabatic connection - higher accuracy at lower cost}

\begin{abstract}
Non-covalent interactions (NCIs) play a crucial role in biology,  chemistry,  material science, and everything in between. To improve pure quantum-chemical simulations of NCIs, we propose a methodology for constructing approximate correlation energies by combining an interpolation along the M{\o}ller adiabatic connection (MP AC) with a regularization and spin-scaling strategy applied to MP2 correlation energies. This combination yields $c_{\rm os}\kappa_{\rm os}$-SPL2,   which exhibits superior accuracy for NCIs compared to any of the individual strategies. With the $N^4$ formal scaling,  $c_{\rm os}\kappa_{\rm os}$-SPL2,  is competitive or often outperforms more expensive dispersion-corrected double hybrids for NCIs.The accuracy of $c_{\rm os}\kappa_{\rm os}$-SPL2 particularly shines for anionic halogen bonded complexes, where it surpasses standard dispersion-corrected DFT by a factor of 3 to 5. 


\end{abstract}
\maketitle

Non-covalant interactions (NCI) play a crucial role in a variety of fields including biology, chemistry, material science, and everything in between\cite{hohenstein12,riley11,lao15,SedJanPitRezPulHob-JCTC-13,alhamdani19,grimme16,dubecky16,christensen16,dilabio13,GAEK10,hobza11,fabiano14,fabiano17}. Second order M{\o}ller-Plesset pertubation theory (MP2) has been used extensively to study NCIs\cite{riley11,hobza88} because it includes dispersion interactions and also has more favorable scalings compared to other wavefunction methods such as CCSD(T).
However,  MP2 does have its downsides because it is known to fail for $\pi$-$\pi$ stacking complexes and generally for NCIs involving highly polarizable molecules\cite{SheLoiRetLeeHea-JPCL-21}. 
More recently,  it has been observed that MP2 and even the whole MP series have increasingly large errors for large complexes\cite{Ngu-Fur-JCTC-2020}, restricting the chemical space of NCIs that MP2 can be meaningfully applied to.

Multiple methods have been introduced to increase the accuracy of MP2 by manipulating the underlying MP2 equations\cite{Gri-JCP-03,JunLocDutHea-JCP-04,LeeHea-JCTC-18,SheLoiRetLeeHea-JPCL-21,LoiBerLeeHea-JCTC-21}, such as spin-component-scaled (SCS) MP2 \cite{Gri-JCP-03}, and spin-opposite-scaled (SOS) MP2\cite{JunLocDutHea-JCP-04}. SOS-MP2 does not only often improve the accuracy of MP2,  but is also cheaper (scales formally as $N^4$) than original MP2 ($N^5$). Another way to improve MP2 energies is to prevent their divergence when
the HOMO-LUMO orbital gap closes through regularization
 \cite{LoiBerLeeHea-JCTC-21,SheLoiRetLeeHea-JPCL-21}, and to mix MP2 with density functional theory (DFT)
as in double hybrids (e.g., B2PLYP)  \cite{Gri-JCP-2006, SchGri-RSC-2007, SSM19, XYG3, XYG7, SonVucSimBur-JPCL-21}. 
However,  (with some exceptions~\cite{XYG3,XYG7,SSM19}), when there is no dispersion correction, 
double hybrids typically worsen MP2 for NCIs \cite{VucGorDelFab-JPCL-18,Gri-JCP-2006,SchGri-RSC-2007,GAEK10,CalEhlHanNeuSpiBanGri-JCP-2019,CalMewEhlGri-PCCP-2020}. 

In our previous work \cite{DaaFabDelGorVuc-JPCL-21}, we introduced a new class of functionals that directly approximate the M{\o}ller Plesset adiabatic connection (AC)\cite{GiaGorDelFab-JCP-18,DaaGroVucMusKooSeiGieGor-JCP-20},  which is the AC that gives an exact expression for the quantum-chemical correlation energy and has the MP series as weak coupling Taylor expansion. 
The idea behind these functionals is that by adding information for the large coupling limit,  we can introduce a curvature in the underlying AC curve that is completely missed by MP2\cite{VucFabGorBur-JCTC-20}.  This curvature has been shown to play a very important role in describing NCIs,  particularly as the ratio between dispersion and electrostatics in NCIs grows.  \cite{VucFabGorBur-JCTC-20}
Our MP AC approach is fundamentally different from double hybrids,  
which rely on the error cancellations between exact exchange and MP2 
and their approximate semilocal counterparts. 
Instead, our MP AC approach uses the exact information from the weak coupling limit of MP AC - the full amounts of 
both exact exchange and MP2. One then recovers correlation energy with MP AC, by performing an interpolation between its weak- and strong-coupling limits. 
The resulting functionals 
based on this interpolation strategy either provide major improvements or are on par with dispersion-enhanced (double) hybrids and massively improve over MP2, particularly for large, $\pi-\pi$-stacking, and charge-transfer complexes~\cite{DaaFabDelGorVuc-JPCL-21}. 
The MP AC interpolation approach introduces a negligible additional cost to MP2 calculations,
and because of that, it can also be viewed  
 as a zero-cost correction to MP2 correlation energies. Consequently, the computational cost of current MP AC functionals is the same as that of double hybrids. Nevertheless, their $N^5$ formal scaling can be prohibitive for large complexes, 
 especially when compared to the more feasible $N^4$ formal scaling of standard DFT hybrids.

\begin{figure}[t]
    \centering
    \includegraphics[width=0.95\linewidth]{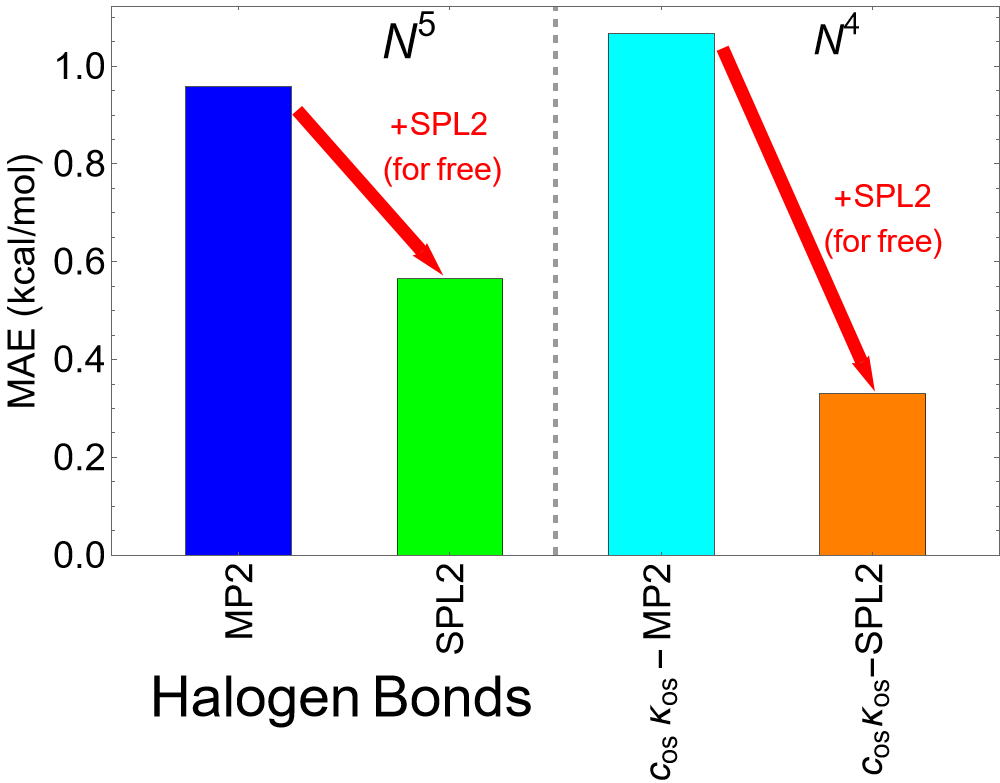}
\caption{The MAE (Mean Absolute Error) of MP2, SPL2, $c_{\rm os}\kappa_{\rm os}$-MP2 and $c_{\rm os}\kappa_{\rm os}$-SPL2 for the halogen bonded complexes of B30\cite{BauAlkFroElg-JCTC-13} and X40\cite{RezRilHob-JCTC-12} (see Fig.~\ref{fig:hal_bonds} for more detail on the complexes and 
Table~\ref{tab:funcs} for individual methods)}
\label{fig:fig0}
\end{figure}
In the present work,  we demonstrate the robustness of our MP AC functionals by applying them to different variants of MP2 and show how their cost and accuracy can be improved simultaneously.
Utilizing this corrective power, we combine the MP AC approach to 
only the spin-opposite part of MP2 and the regularized version thereof,
yielding massive improvements over SOS-MP2 for NCIs. This scaling-lowering strategy ($N^5 \to N^4$) not only retains the accuracy of the original MP AC approach that uses 'bare' MP2, but also gives major improvements across various classes of NCIs. 

Fig.~\ref{fig:fig0} demonstrates the corrective power of a 
successful
MP AC interpolation form, called SPL2\cite{DaaFabDelGorVuc-JPCL-21},  i.e. containing two terms of the Seidl, Perdew, and Levy (SPL)\cite{SeiPerLev-PRA-99} interpolation formula. The individual bars show mean absolute errors (MAE) for the interaction energies of halogen-bonded complexes. 
When "bare" MP2 (blue) is applied to these complexes, we get an MAE of slightly less than 1 kcal/mol. Applying SPL2 (the red arrow) to it the error reduced to less than 0.6 kcal/mol. This is a substantial change relative to the magnitude of interaction energies of halogen-bonded complexes. 
Now we consider a cost-lowering strategy and we take only the spin-opposite part of it, re-scale 
it and apply to it a specific regularization \cite{LoiBerLeeHea-JCTC-21,SheLoiRetLeeHea-JPCL-21} to obtain $c_{\rm os}\kappa_{\rm os}$-MP2, which will be described in detail in the following. Once we go from MP2 (blue; $N^5$ scaling) to (light blue; $N^4$ scaling) the MAE increases slightly. However, SPL2 corrects $c_{\rm os}\kappa_{\rm os}$-MP2 to yield $c_{\rm os}\kappa_{\rm os}$-SPL2 (orange), which gives now improved results (MAE less than 0.4 kcal/mol) even w.r.t. more expensive and original SPL2 method (green). 
At this point, we should remember that both of the red arrows in Fig.~\ref{fig:fig0} practically come at no additional cost to the original MP2 calculations, be it MP2 or $c_{\rm os}\kappa_{\rm os}$-MP2. 
In what follows, we will detail the theoretical foundation for the construction of our methods. 

 {\it Constructing MP AC.-} The Møller-Plesset adiabatic connection (MP AC) connects the (non-interacting) Hartree-Fock (HF) system to the physical system through the following Hamiltonian\cite{Per-IJQC-18,DaaGroVucMusKooSeiGieGor-JCP-20,SeiGiaVucFabGor-JCP-18,DaaKooGroSeiGor-JCTC-22},
\begin{equation}\label{eq:HlambdaHF}
	\hat{H}_{\lambda}=\hat{T}+\hat{V}_{\rm ext}+\lambda \hat{V}_{ee}+
	\Big(1-\lambda \Big)
	\Big(\hat{J}+\hat{K} \Big),
\end{equation}
with $\lambda \geq 0$ the coupling constant, $\hat{T}$ the kinetic energy, $\hat{V}_{ee}$ the electron repulsion operator, $\hat{V}_{\rm ext}$ representing the external potential, and $\hat{J}=\hat{J}[\rho^{\rm HF}]$ and $\hat{K}=\hat{K}[\{\phi_i^{\rm HF}\}]$ being the standard HF Coulomb and exchange operators. The last two are functionals of the HF density, $\rho^{\rm HF}$, and the occupied orbitals, $\phi_i^{\rm HF}$, respectively and are $\lambda$-independent. 
The correlation energy (exact energy minus HF energy) can be obtained by applying the Hellman-Feynman theorem to Eq.~\eqref{eq:HlambdaHF},\cite{Ngu-Fur-JCTC-2020,Per-IJQC-18,BurMarDaaGorLoo-JCP-21,SeiGiaVucFabGor-JCP-18,DaaGroVucMusKooSeiGieGor-JCP-20,DaaFabDelGorVuc-JPCL-21,DaaKooGroSeiGor-JCTC-22}
\beq \label{eq:ac}
E_{\rm c}= \int_0^1 W_{\rm c,\lambda} \dd \lambda,
\eeq
where $W_{\rm c,\lambda}$ is the AC integrand,
\beq \label{eq:wchf}
W_{c, \lambda}=\langle \Psi_\lambda
|\hat{V}_{\rm ee} - \hat{J}-\hat{K}|
\Psi_\lambda \rangle 
- \langle  \Psi_0
 |\hat{V}_{\rm ee} - \hat{J}-\hat{K}
 |  \Psi_0\rangle,
\eeq
with $\Psi_\lambda$ being the wavefunction that minimizes $\hat{H}_{\lambda}$. 
The small-$\lambda$ expansion of $W_{c,\lambda}$ is then the standard MP series,
\begin{equation}\label{eq:WHFMP}
W_{c,\lambda\rightarrow 0}=\sum_{n=2}^\infty n\,E^{{\rm MP}n}_{c}\,\lambda^{n-1}.
\end{equation}
The large $\lambda$ expansion of the MP AC has the form,\cite{SeiGiaVucFabGor-JCP-18,DaaGroVucMusKooSeiGieGor-JCP-20}
\begin{equation}  \label{eq:large}
W_{c,\lambda\rightarrow\infty} = W_{c,\infty} + \frac{W_{\frac{1}{2}}}{\sqrt{\lambda}}+\frac{W_{\frac{3}{4}}}{\lambda^{\frac{3}{4}}}+\dots
\end{equation}
with 
\begin{equation}
 W_{c,\infty}  = E_{el}[\rho^{\rm HF}]+E_x \label{eq:Wcinffinal},
\end{equation}
where $E_{\rm el}[\rho]$ is\cite{SeiGiaVucFabGor-JCP-18,DaaGroVucMusKooSeiGieGor-JCP-20} the classical electrostatic energy of $N$ point charges bound by a charge density $\rho({\bf r})$ of opposite sign, with $\int \rho({\bf r}) d{\bf r}=N$.  While gradient expansion approximations have recently been introduced for both $E_{\rm el}$ and $W_{\frac{1}{2}}$,\cite{DaaKooGroSeiGor-JCTC-22}  here we will approximate $W_{c,\infty}$ from the following inequality,
\begin{equation}
W_{c,\infty}\leq   W^{\rm DFT}_{\infty}+E_{x}.
\end{equation}
This inequality links $W_{c,\infty}$ to $W^{\rm DFT}_{\infty}$ from the strong coupling limit of the DFT density-fixed AC.\cite{SeiGorSav-PRA-07,GorVigSei-JCTC-09,GorSeiSav-PCCP-08,GorSeiVig-PRL-09,GroKooGieSeiCohMorGor-JCTC-17,VucGerDaaBahFriGor-WIR-22} 
In our previous work,\cite{DaaFabDelGorVuc-JPCL-21}  we introduced $W^{\alpha,\beta}_{c,\infty}$ based on this inequality as
\begin{equation}\label{eq:DFTMPC}
W^{\alpha,\beta}_{c,\infty}=  \alpha W^{\rm DFT}_{\infty}+\beta E_{x},
\end{equation}
so that $\alpha$ and $\beta$ can be used as fitting parameters. This has been done to approximate the effect of $W_{\frac{1}{2}}$ on the MP AC without including it directly. 
For $W^{\rm DFT}_{\infty}$ in Eq.~\eqref{eq:DFTMPC},  we use the accurate PC model approximation,\cite{SeiPerKur-PRA-00,SeiGorSav-PRA-07,MirSeiGor-JCTC-12}
\begin{align}
\label{eq:pc}
W_{\infty}^{\rm DFT} [\rho ^{\rm HF}]\approx \int \left[A\rho^{\rm HF}(\mathbf{r})^{4/3} + B\frac{|\nabla \rho^{\rm HF}(\mathbf{r})|^2}{\rho^{\rm HF}(\mathbf{r})^{4/3}}\right]\mathrm{d}\mathbf{r},
\end{align}
where $A=-1.451$, $B=5.317\times10^{-3}$.
In our previous work, two new MP AC functionals,\cite{DaaFabDelGorVuc-JPCL-21} SPL2 and MPACF-1, have been introduced. The SPL2 model for $W_{c,\lambda}$ reads 
\beq \label {eq:spl2-M}
 W_{c,\lambda}^{\rm SPL2} = C_1 -  \frac{m_1}{\sqrt{1 + b_1 \lambda} } - \frac{m_2}{\sqrt{1 + b_2 \lambda} },
\eeq
with $b_1$, $m_1$ and $C_1$ fixed by the exact constraints: (i) $W_{c,\lambda}^{\rm SPL2}$ vanishes at 0; (ii) its  derivative at $\lambda=0$ is equal to $2 E_c^{\rm MP2}$ (Eq.~\ref{eq:WHFMP}); (iii) it converges to  $W_{c,\infty}$ at $\lambda \to \infty$ (Eq.~\ref{eq:large}).
The remaining parameters are fitted to the S22 set of NCIs\cite{JurSpoCerHob-PCCP-06}. 
MPACF1 (hereinafter F1) approximates $E_c$ directly,  resulting in,
\begin{equation}
E_{\rm c,\lambda}^{\rm F1}=-g \lambda +\frac{g (h+1) \lambda}{\sqrt{d_1^2 \lambda +1}+h \sqrt[4]{d_2^4 \lambda+1}},
\end{equation}
with $g$ and $h$ given in Ref. \onlinecite{DaaFabDelGorVuc-JPCL-21}, 
and $E_{\rm c,\lambda=1}=E_c$. In the original F1 formulation, $\alpha$ and $\beta$ were set to 1 to reproduce the HEG correctly.
However, here we use the general F1-form where [$d_1$,$d_2$,$\alpha$,$\beta$] were fitted to the S22 dataset. 
The advantage of F1 over SPL2 is that the former has the $\lambda^{-3/4}$ term in the model $W_{c,\lambda\to\infty}$, which is missed by SPL2. Although the SPL2 and F1 functionals are not size-consistent by themselves, we recover size-consistency at no extra computational by applying the correction of ref.~\citenum{VucGorDelFab-JPCL-18}.

 {\it \bf Different MP2 flavors yield different MP ACs.-} 

Here we consider the following general form for spin-scaled and regularized MP2\cite{SheLoiRetLeeHea-JPCL-21} correlation energy:
 \begin{equation}
    E_{c}=c_{\rm os}E_{c}^{\kappa_{\rm os}-\text{MP2}} + c_{\rm ss}E_{c}^{\kappa_{\rm ss}-\text{MP2}},
\label{eq:mp24p}
\end{equation}
with  the opposite-spin (os) and the same-spin (ss) parts given by (closed-shell systems are assumed throughout this work),
\begin{equation}
    E_{c}^{\kappa_{\rm os}-\text{MP2}}=-\sum_{abrs} \frac{\langle ab|rs\rangle^2}{\Delta_{ab}^{rs}}(1-e ^{-\kappa_{\rm os}\Delta_{ab}^{rs}})^2
\end{equation}
and
\begin{equation}
    E_{c}^{\kappa_{\rm ss}-\text{MP2}}=-\sum_{abrs} \frac{\langle ab|rs\rangle[\langle ab|rs\rangle-\langle rs|ba\rangle]}{\Delta_{ab}^{rs}}(1-e ^{-\kappa_{\rm ss}\Delta_{ab}^{rs}})^2
\end{equation}
with $\Delta_{ab}^{rs} = \epsilon_r + \epsilon_s - \epsilon_a -\epsilon_b$ being an orbital gap between a pair of occupied ($r$ and $s$) and a pair of virtual orbitals ($a$ and $b$).

When the $\kappa$ parameters are set to $\infty$, 
 Eq. \eqref{eq:mp24p} reduces to  the SCS-MP2 scheme \cite{Gri-JCP-03}, with the two parameters $c_{os}$ and $c_{ss}$ (when these are further set to $1$,  Eq. \eqref{eq:mp24p} reduces to  the standard MP2). 
 The os part formally scales as $N^4$ whereas the ss part as $N^5$.  
Setting $c_{\rm ss}=0$ defines the cheaper scaled opposite-spin-only (SOS)  variant (with, e.g., $c_{\rm os}=1.3$)~\cite{JunLocDutHea-JCP-04}.
When the $\kappa$ parameters are set to a finite value the divergence
for vanishing gap is prevented.
For NCI, $\kappa$-MP2 provides significant improvements over MP2, 
particularly for values $\kappa \sim 1.2$. \cite{SheLoiRetLeeHea-JPCL-21,RetSheLeeHea-JCTC-22,SanMar-JPCL-22}

\begin{table}[t]
\caption{The 6 functionals that we have studied in this work (in addition to MP2 and SPL2) and the parameters that were fitted to the S22 dataset. 
The parameters for other 14 combinations as well as the corresponding MP AC ones be found in Table S1 in the SI.}
\begin{tabular}{l||rrrrr}
   Method        & $c_{ss}$ & $c_{os}$ &  $\kappa_{\rm ss}$ & $\kappa_{\rm os}$   \\ \hline\hline
                                                 
MP2                                   & 1 & 1& $\infty$ & $\infty$ \\
SPL2                                  &  1 & 1 &  $\infty$ & $\infty$  \\
\hline
$\kappa_{\rm ss},\kappa_{\rm os}$-MP2 &  1 & 1 & 0.9 & 1.4 \\
$\kappa_{\rm ss},\kappa_{\rm os}$-SPL2 & 1& 1 &  1.1 & 1.7 \\
$\kappa$-F1                            & 1 &1 & 1.5 & 1.5 \\
\hline
$c_{\rm os}\kappa_{\rm os}$-MP2   & 0 & 2.1 &  0 & 0.9  \\
$c_{\rm os}\kappa_{\rm os}$-SPL2   & 0 & 2.1 & 0 & 1.3  \\
$c_{\rm os}$-SPL2                  & 0 & 1.8 & $\infty$ & $\infty$ \\   
\end{tabular}
\label{tab:funcs}
\end{table}


In the general expression of Eq.\eqref{eq:mp24p}, we initially have four parameters that can be fitted. However, here we consider an OS variant with two parameters that require fitting, namely $c_{os}$ and $\kappa_{\rm os}$, while $c_{ss}$ is preset to 0. Our second MP2 variant has both ss and os spin channels, but still, two parameters to be fitted ($\kappa_{\rm ss}$ and $\kappa_{\rm os}$) with $c_{ss}$ and $c_{os}$ preset to 1.
To each of those, we can add the MP AC correction on top, yielding a large number of different functionals for the correlation energy. 
Here we focus on the three that are most promising (Tab.~\ref{tab:funcs}), and others can be found in Tab. S1 of the SI.   
For naming MP AC methods, we use the following notation:
The string in front of the functional stands for which version of MP2 is used in that functional. 
So if regular MP2 is used within MP AC, the functional will just be called SPL2 or F1. 
If a modified version of MP2 has been used, such as $c_{\rm os},\kappa_{\rm os}$-MP2, then the functional is called $c_{\rm os},\kappa_{\rm os}$-SPL2 or $c_{\rm os},\kappa_{\rm os}$-F1. 
Finally,  $\kappa_{\rm ss} \kappa_{\rm os}$-SPL2
 means that SPL2 interpolation has been used 
 with $\kappa_{\rm ss} \kappa_{\rm os}$-MP2 correlation energy (Eq.~\ref{eq:mp24p} with $c_{\rm ss} = c_{\rm os} =1$).

We acknowledge that employing a modified MP2 correlation energy as half of the initial slope of $W_{c,\lambda}$ deviates from the exact MP AC theory, which dictates the initial slope to be the unaltered MP2. 
Nevertheless, we adopt this pragmatic approach, 
recognizing the inherent flexibility of our MP AC construction, reflected in the endpoint of our MP AC curve, $W^{\alpha,\beta}_{c,\infty}$, which can adjust itself to the modified initial slope through the $\alpha$ and $\beta$ parameters. Moreover, for larger NCI complexes, the radius of convergence of the MP 
series could be exceptionally small.~\cite{Ngu-Fur-JCTC-2020,NguHerFloFur-ES-22} 
This could reduce the utility of the actual slope of the MP AC for approximating the true MP AC. 
Therefore, our approach, which involves regularization and scaling modifications of the MP AC initial slope, 
offers us a heuristic to capture MP ACs pertaining to NCIs. 
Simultaneously, this method provides an opportunity to decrease the computational cost of our MP AC construction, for instance, by using only the os part of MP2.

Returning to Tab.\ref{tab:funcs}, we distinguish methods with computational scaling of $N^5$ (upper part of the table) from those with $N^4$ (lower part). 
We will compare the accuracy of these functionals with other common functional
B2PLYP-D3\cite{Gri-JCC-2006,GAEK10, ElsHobFraKax-JCP-01,WuYan-JCP-02,Gri-JCC-04,BecJoh-JCP-07,GoeHanBauEhrNajGri-RSC-2017} (similar cost to our $N^5$ methods) 
and B3LYP-D3. Note that B3LYP-D3 has the same scaling as our $N^4$ methods, 
but likely has a smaller prefactor.
Nevertheless, SOS-MP2 can be implemented very efficiently~\cite{ForFraLenVis-JCTC-20,GAEK10, ElsHobFraKax-JCP-01,WuYan-JCP-02,Gri-JCC-04,BecJoh-JCP-07,GoeHanBauEhrNajGri-RSC-2017}),
with its wall times often being shorter than those for the HF calculation~\cite{ForFraLenVis-JCTC-20}."

\begin{figure*}[t]
\centering
\includegraphics[width=.95\linewidth]{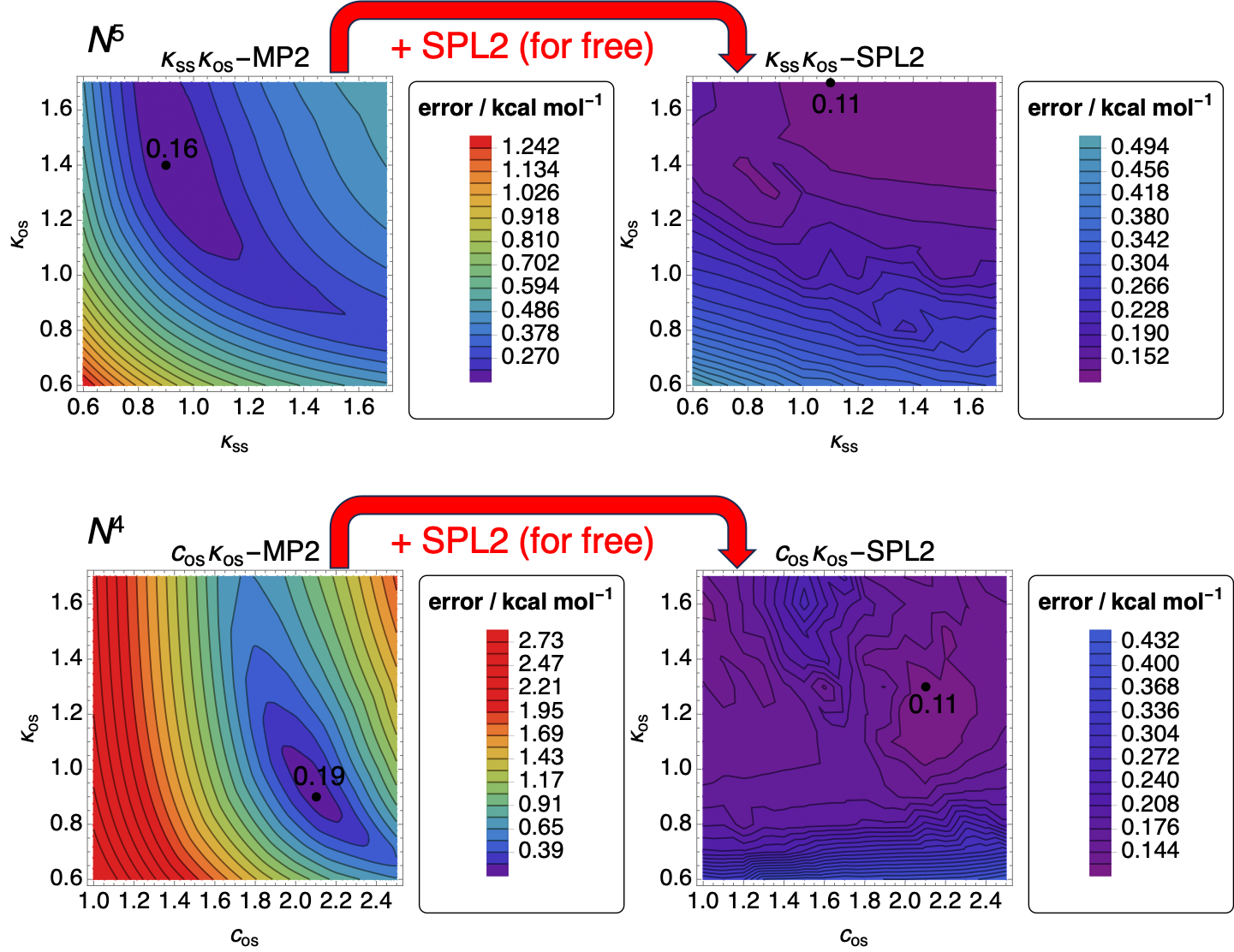}
\caption{Top: The error between the interacting energy for the S22 dataset between the CCSD(T) reference data and $\kappa_{\rm ss}\kappa_{\rm os}$-MP2 and $\kappa_{\rm ss}\kappa_{\rm os}$-SPL2 for a range of $\kappa$'s between 0.6 and 1.7.
Bottom: The same but for $c_{\rm os}\kappa_{\rm os}$-MP2 and $c_{\rm os}\kappa_{\rm os}$-SPL2 for a range of $c_{\rm os}$ values from 1 to 2.5 and $\kappa_{\rm os}$'s between 0.6 and 1.7. The black dots denote the minimum energy of the contour plot with its corresponding value. A figure containing the curves of $\kappa$-MP2 and $\kappa$-SPL2, so the diagonals of the two top panels, 
can be found in the SI (figure S2)}
\label{fig:ksskoscos}
\end{figure*}

 {\it \bf Training of our functionals on S22.-} Similar to previous work,~\cite{DaaFabDelGorVuc-JPCL-21} 
 we train the empirical parameters in our functional
 by minimizing their MAEs for the S22 dataset.\cite{JurSpoCerHob-PCCP-06}
 Notice that for each $\kappa$ value,  the MP2 correlation energy needs 
 to be re-calculated and we consider here $\kappa$-values between 0.6 and 1.7 with a 0.1 step.
We also use $c_{\rm os}$ values
from $1.0$ to $2.5$ with the step of $0.1$.
Note that the re-optimization of the MP AC parameters 
is needed any time an alteration to MP2 is made. 
Thus, 
for each combination of $\kappa_{\rm os}$ and $\kappa_{\rm ss}$ (or $c_{\rm os}$), 
the corresponding 
parameters of the functionals are given in Tab.~S2 in the SI. 

In Figure~\ref{fig:ksskoscos}, we show contour plots with the MAE for the S22 dataset with the parameters of our selected methods (functionals). 
The two left-side panels display the S22 MAE contour plots for two distinct MP2 versions ($\kappa_{\rm ss}\kappa_{\rm os}$-MP2, an $N^5$ method; and $c_{\rm os}\kappa_{\rm os}$-MP2 an $N^4$ method) against their respective parameters. 
Conversely, the two right-side panels show the same plots after applying the SPL2 correction.

Note that the SPL2 contour plots on the right are the result of re-optimizing the four SPL2 parameters ($b_2$, $m_2$, $\alpha$, $\beta$) for each combination of the $\kappa_{\rm ss}$ and $\kappa_{\rm os}$ parameters (upper panel) and $c_{\rm os}$ and $\kappa_{\rm os}$ parameters (lower panel).
The transition from the uncorrected MP2 versions (left panels) to the SPL2 corrected versions (right panels) yields a narrower range of MAEs, 
echoing our previous discussion on the adaptability of our MP AC construction to different MP2 inputs. 
The robustness of the SPL2 correction, regardless of the MP2 input, 
is further evidenced by the significant areas within both SPL2 contour plots that exhibit minimal MAE variation. 

The top-left panel focuses on the $\kappa_{\rm ss}\kappa_{\rm os}$-MP2 
with regularizers $\kappa_{\rm ss}$ and $\kappa_{\rm os}$ as parameters. 
Here, the MAE is considerably more sensitive to the $\kappa_{\rm os}$ regularizer than to its same-spin counterpart. 
The minimum MAE is achieved at $\kappa_{\rm os} = 1.4 $ and $\kappa_{\rm ss} = 0.9 $ (0.16 kcal/mol). 
If we restrict $\kappa = \kappa_{\rm ss} = \kappa_{\rm os} $ as done 
in previous work~\cite{SheLoiRetLeeHea-JPCL-21}, 
the minimum is at $\kappa = 1.1$ with an MAE of 0.21 kcal/mol (see Fig.~S2 in the SI). 
Adding the SPL2 correction to $\kappa_{\rm ss}\kappa_{\rm os}$-MP2 (moving from the top-left to top-right panel) 
shifts the minimum to $\kappa_{\rm os} = 1.1 $ and $\kappa_{\rm ss} = 1.7 $, reducing it to 0.11 kcal/mol. 
While the reduction is modest, it is likely approaching the accuracy limit of MP2-like methods for the S22 dataset. 
Importantly, we observe a relatively flat area around the minimum, 
suggesting that the MAE is almost constant as long as $\kappa_{\rm ss}$ 
and $\kappa_{\rm os}$ exceed approximately 1.1.

The lower-left panel illustrates MAE for the opposite-spin-only $c_{\rm os}\kappa_{\rm os}$-MP2 with its parameters. 
Now the MAE depends substantially more on the $c_{\rm os}$ scaling factor than on the $\kappa_{\rm os}$ regularizer. 
The minimum is achieved at $c_{\rm os}=2.1$ and $\kappa_{\rm os} = 0.9$, 
resulting in a MAE of 0.19 kcal/mol. 
This value is only marginally higher than the one from the upper panel, 
despite a significant reduction in computational cost. With the addition of the SPL2 correction to $c_{\rm os}\kappa_{\rm os}$-MP2 (transition from lower-left to lower-right panel), 
the minimum MAE dips to 0.11 kcal/mol, slightly shifted to $c_{\rm os}=2.1$ and $\kappa_{\rm os} = 1.3$. 
This MAE yields a significant improvement over $c_{\rm os}\kappa_{\rm os}$-MP2, and is on par with the more expensive
$\kappa_{\rm ss}\kappa_{\rm os}$-SPL2 (top-right panel). 
Thus, through these explorations, we see the robustness of SPL2 
to correct various MP2 inputs.

\begin{table}[t]
    \centering
    \begin{tabular}{c||ccccc}
\text{Method/MAE} & \text{S22} & \text{NCCE31} & \text{S66x8} &\text{X40} & \text{B30}  \\\hline \hline
 \text{MP2} & 0.86 & 0.5 & 0.39 & 0.34 & 0.86 \\
 \text{SPL2} & 0.15 & 0.25 & 0.15 & 0.16 & 0.48\\
 \text{$\kappa $-F1} & 0.14 & 0.32 & 0.13 & 0.17 & 0.7 \\
 \text{B2PLYP-D3} & 0.15\cite{GoeHanBauEhrNajGri-RSC-2017} & 0.61\cite{LiaNee-JCTC-15} & \textbf{0.11}\cite{BraKesKozMar-PCCP-16} & 0.18 & 0.72 \\\hline
 $c_{\text{\rm os}}\kappa _{\text{\rm os}}-\text{MP2}$ & 0.19 & 0.4 & 0.2 & 0.22 & 0.93\\
 $c_{\text{\rm os}}\kappa _{\text{\rm os}}-\text{SPL2}$ & \textbf{0.11} & 0.27 & \textbf{0.11} & \textbf{0.10}& \textbf{0.45}  \\
$c_{\text{\rm os}}-\text{SPL2}$ & 0.15 & \textbf{0.21} & 0.13 & 0.13 & 0.9 \\
 \text{B3LYP-D3} & 0.31\cite{GoeHanBauEhrNajGri-RSC-2017} & 0.60\cite{LiaNee-JCTC-15} & 0.15\cite{BraKesKozMar-PCCP-16} & 0.22\cite{MarHea-MP-17} & 1.9\cite{MarHea-MP-17} \\
     \end{tabular}
    \caption{The total MAE's of the 5 NCI datasets considered in this work for the 8 studied functionals. Best result for each column is highlighted in boldface.}
    \label{tab:TotMAE}
\end{table}

\begin{figure}[t]
\centering
\includegraphics[width=.95\linewidth]{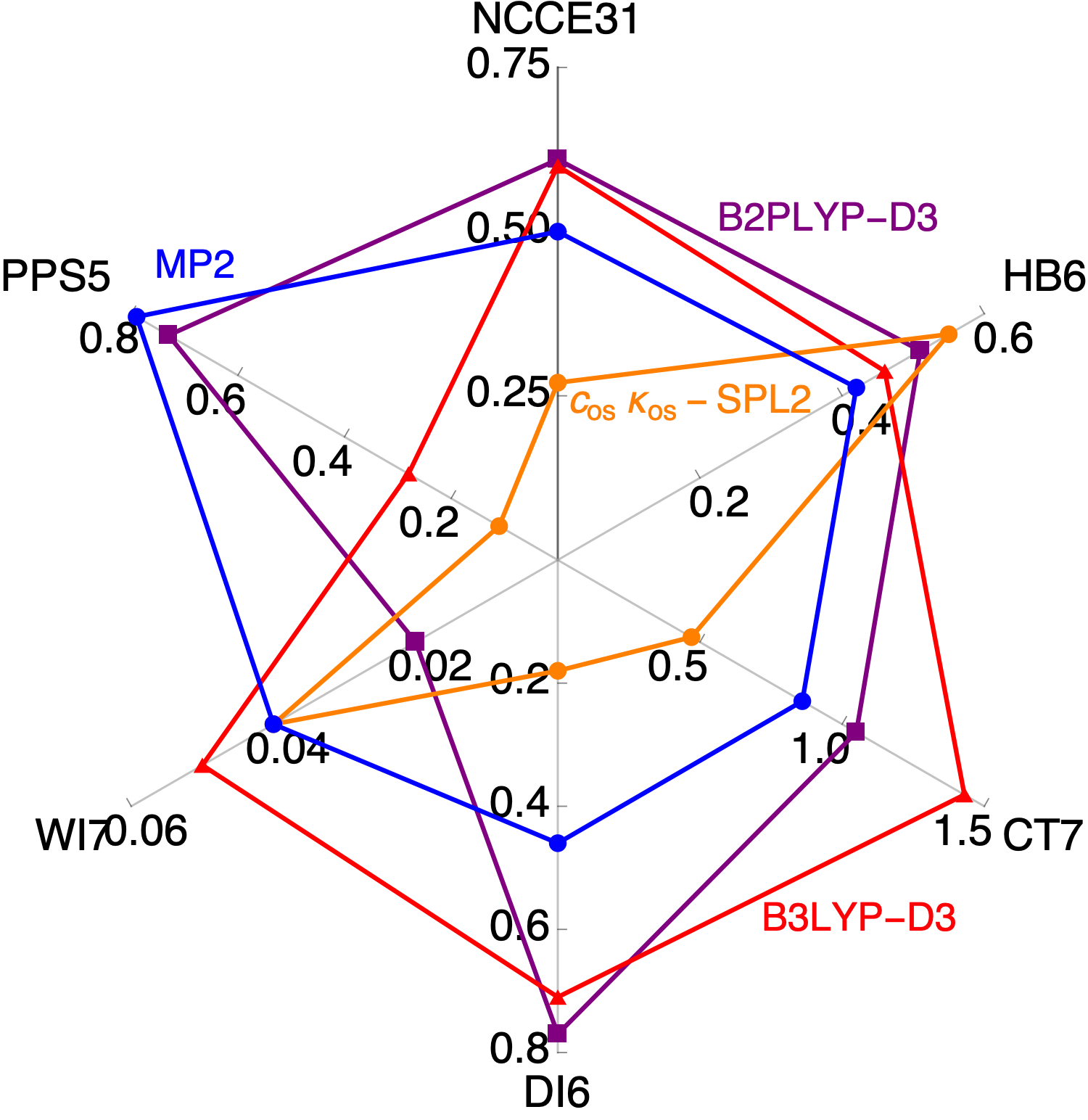}
\caption{A radar plot for NCCE31 containing the MAEs in kcal/mol of the 5 subgroups (HB6, CT7, DI6, WI7 and PPS5) as well as the total MAE of MP2, $c_{\rm os}\kappa_{\rm os}$-SPL2, B3LYP-D3 and B2PLYP-D3. A table containing the MAEs of these methods as well as the other 4 methods studied in the paper can be found in the SI (Tab. S3). HB stands for hydrogen bonds, 
CT for charge transfer interaction,
DI for dipole interactions,
WI for weak interactions,
and PPS for $\pi$-$\pi$-stacking interactions}
\label{fig:radarplot}
\end{figure}

{\it \bf $c_{\rm os}\kappa_{\rm os}$-MP2 as our workhorse
-} After training our functionals on S22, we test them to a variety of NCI datasets, with the results shown in Table~\ref{tab:TotMAE}. Alongside S22, we include NCCE31~\cite{ZhaTru-JPCA-05, ZhaXuJunGod-PNAS-2011, PevTru-PTRSA-14}
, S66x8, X40, and B30. 
From Table~\ref{tab:TotMAE}, we can see that
$c_{\rm os}\kappa_{\rm os}$-MP2 has the overall best performance (even when compared to more expensive $N^5$ methods). 

In Figure~\ref{fig:radarplot}, we use radar plots to show the performance of various functionals across five interaction-specific subsets in the NCCE31 dataset, including hydrogen bonds (HB6), charge-transfer interactions (CT7), dipole interactions (DI6), weak interactions (WI7), and $\pi$-$\pi$-stacking interactions (PPS5). 
We can see that $c_{\rm os}\kappa_{\rm os}$-SPL2 exhibits superior performance in CT7, DI6, and PPS5 subsets, as well as the NCCE31 dataset as a whole. 
While B2PLYP-D3 outperforms $c_{\rm os}\kappa_{\rm os}$-SPL2 for the WI7 subset, the differences in MAEs are very small. In the HB6 subset,  MP2 is the top performer, but the accuracy of $c_{\rm os}\kappa_{\rm os}$-SPL2 is still satisfactory and is similar to that of B2PLYP-D3.

\begin{table}[t]
    \centering
    \begin{tabular}{c||c|ccc}
 \text{Method} & \text{MAE} & \text{H-bonds} & \text{Dispersion} & \text{Mixed} \\\hline\hline
 \text{MP2} & 0.39 & 0.14 & 0.67 & 0.34 \\
 \text{SPL2} & 0.15 & 0.12 & 0.21 & 0.1 \\
 $\kappa -\text{F1}$ & 0.13 & 0.09 & 0.19 & 0.11 \\
 \text{B2PLYP-D3} & \textbf{0.11} & 0.09 & 0.14 & \textbf{0.08} \\\hline
 $c_{\text{\rm os}}\kappa _{\text{\rm os}}-\text{MP2}$ & 0.2 & 0.14 & 0.32 & 0.14 \\
 $c_{\text{\rm os}}\kappa _{\text{\rm os}}-\text{SPL2}$ & \textbf{0.11} & \textbf{0.08} & 0.16 & 0.1 \\
  $c_{\text{\rm os}}-\text{SPL2}$ & 0.13 & 0.11 & 0.16 & 0.1 \\
 \text{B3LYP-D3} & 0.15 & 0.22 & \textbf{0.13} & \textbf{0.08} \\
  \end{tabular}
    \caption{The total MAE's of the S66x8 dataset of 8 functionals as well as the MAE's for only the hydrogen bonding complexes, the dispersion interactions dominated complexes and the mixed set. Best result for each column is highlighted in boldface.}
    \label{tab:S66x8}
\end{table}

\begin{figure*}[t]
\begin{subfigure}{.49\textwidth}
    \centering
    \includegraphics[width=.95\linewidth]{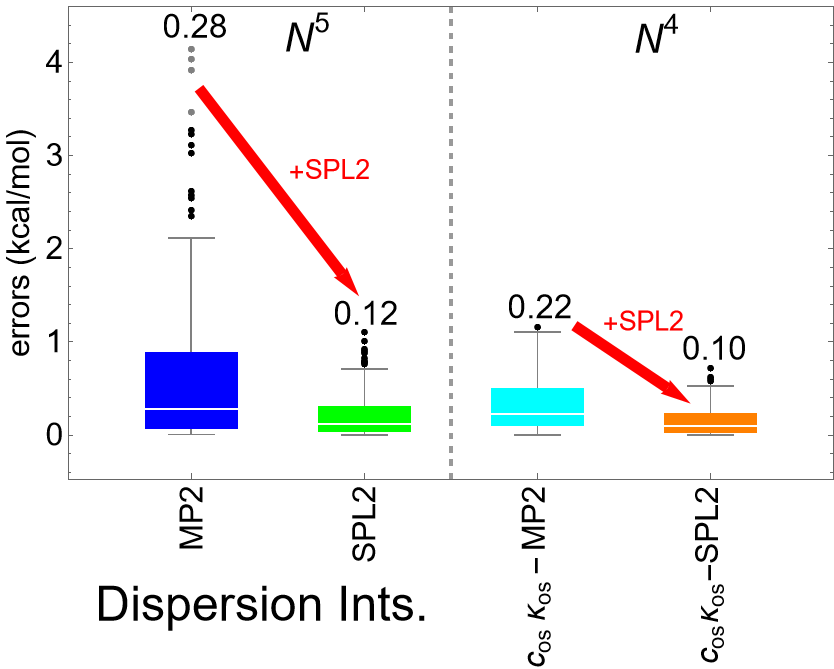}
\end{subfigure}
\begin{subfigure}{.49\textwidth}
    \centering
    \includegraphics[width=.95\linewidth]{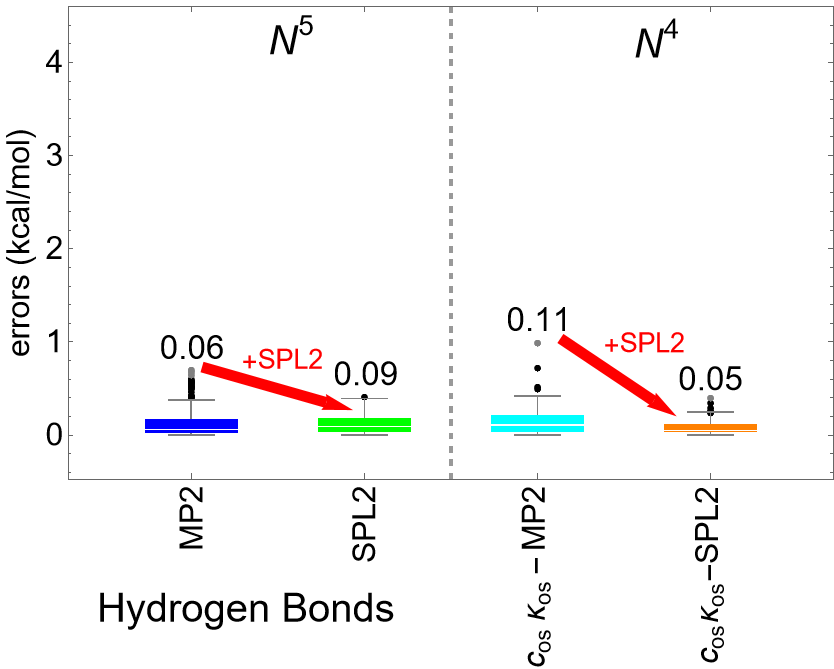}
\end{subfigure}
\begin{subfigure}{.49\textwidth}
    \centering
    \includegraphics[width=.95\linewidth]{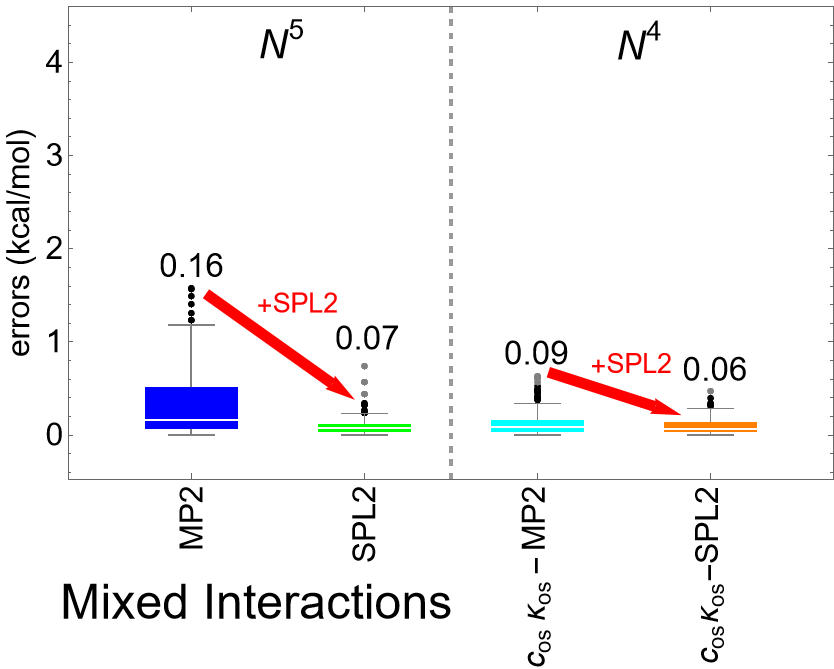}
\end{subfigure}
\begin{subfigure}{.49\textwidth}
    \centering
    \includegraphics[width=.95\linewidth]{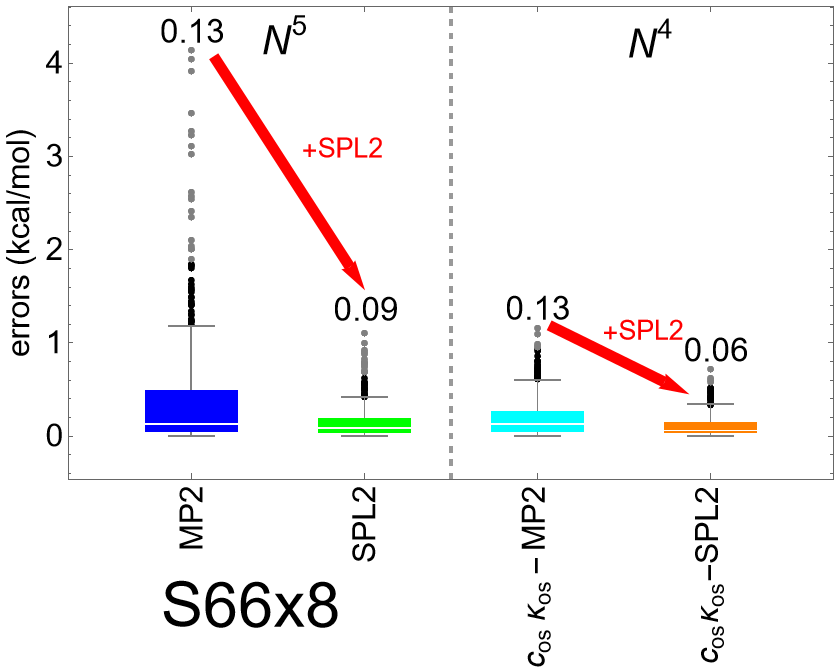}
\end{subfigure}
\caption{The box-plots including outliers as dots containing the errors of MP2, SPL2 and $c_{\rm os}\kappa_{\rm os}$-MP2 and $c_{\rm os}\kappa_{\rm os}$-MP2 for respectively the dispersion interactions(top left), hydrogen bonds (top right), mixed interactions (bottom left) and all interactions in total (bottom right), which show that the SPL2 form corrects the base MP2 form for the different kind of interactions of S66x8. Medians are denoted by a thin white line inside of each box and 
the underlying median values are also given above each box. The boxplots of the remaining functionals can be found in 
Figure~S3 in the SI.
Boxplots showing how errors of different methods vary with the distances between fragments 
can be found in
Figure~S4 in the SI.}
\label{fig:Box_MAE}
\end{figure*}

 {\it \bf S66x8 - accuracy beyond equilibrium.-} Turning our attention to the S66x8 dataset\cite{RezRilHob-JCTC-11, BraKesKozMar-PCCP-16, SanSemMehKarMar-PCCP-22}, we assess the performance of our functionals across NCIs beyond their geometric equilibrium (Tab.~\ref{tab:S66x8}). 
 We can see from the table that $c_{\rm os}\kappa_{\rm os}$-SPL2 ($N^4$ scaling) and B2PLYP-D3 ($N^5$ scaling) deliver twice the accuracy of MP2. For 
 dispersion and mixed interactions, B3LYP-D3 shows a slightly better performance 
 than our $c_{\rm os}\kappa_{\rm os}$-SPL2, but for the hydrogen-bonded complexes
 B3LYP-D3 yields the poorest results.

 From the boxplots shown in Figure~\ref{fig:Box_MAE}, we observe the impact of SPL2 addition on the error distributions of MP2 and $c_{\rm os}\kappa_{\rm os}$-MP2 across individual categories within the S66x8 dataset.
 As expected,\cite{DaaFabDelGorVuc-JPCL-21} SPL2 considerably improves ``bare'' MP2 across all categories. 
 Using SPL2 to correct $c_{\rm os}\kappa_{\rm os}$-MP2 can be considered more challenging as $c_{\rm os}\kappa_{\rm os}$-MP2 already gives significant improvements over MP2.
Nevertheless, SPL2 does also a very good job here, as $c_{\rm os}\kappa_{\rm os}$-SPL2 surpasses $c_{\rm os}\kappa_{\rm os}$-MP2, both in terms of median values and the range of errors across all  S66x8 categories.
Moreover, as demonstrated in Figure~\ref{fig:Box_MAE}, $c_{\rm os}\kappa_{\rm os}$-SPL2 (orange) significantly outperforms the original SPL2 (green), highlighting the dual advantage of our newly developed SPL2 formulation - it provides superior accuracy at a reduced computational cost.

\begin{figure*}[t]
\begin{subfigure}{.49\textwidth}
    \centering
    \includegraphics[width=.95\linewidth]{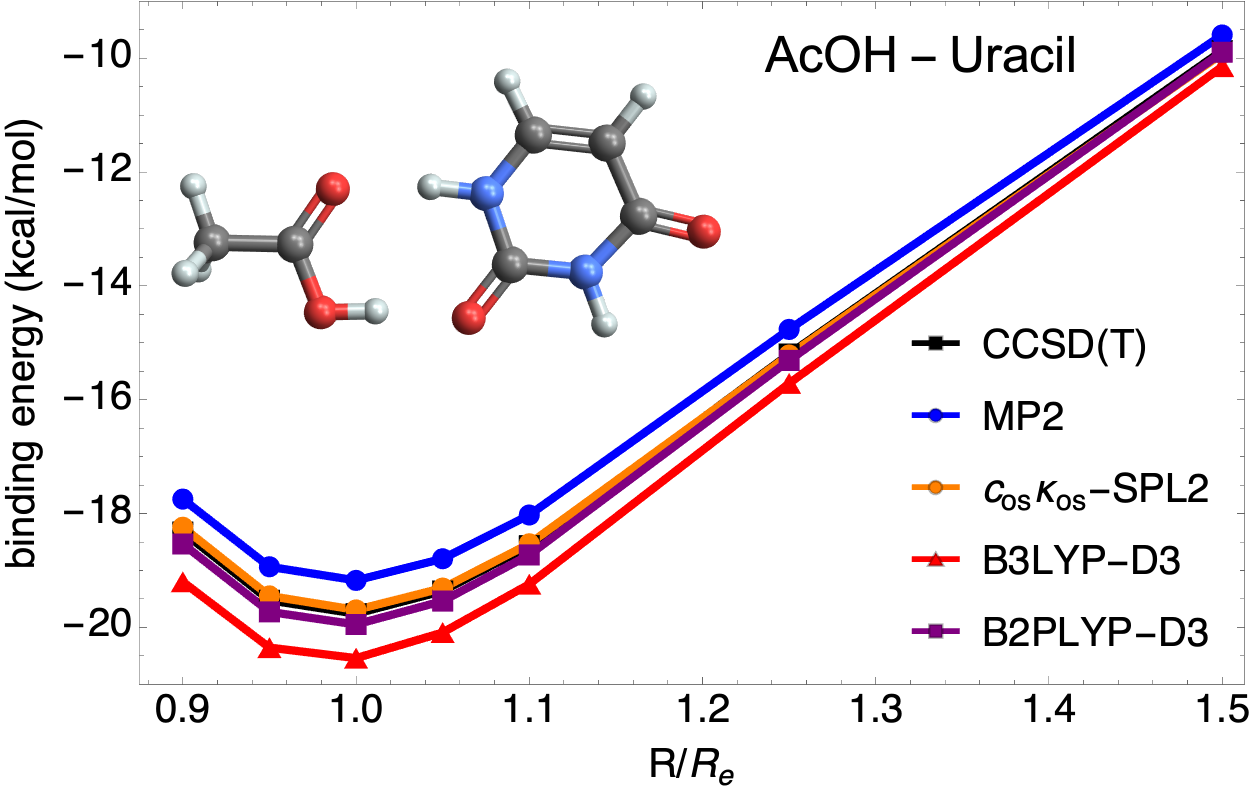}
\end{subfigure}
\begin{subfigure}{.49\textwidth}
    \centering
    \includegraphics[width=.95\linewidth]{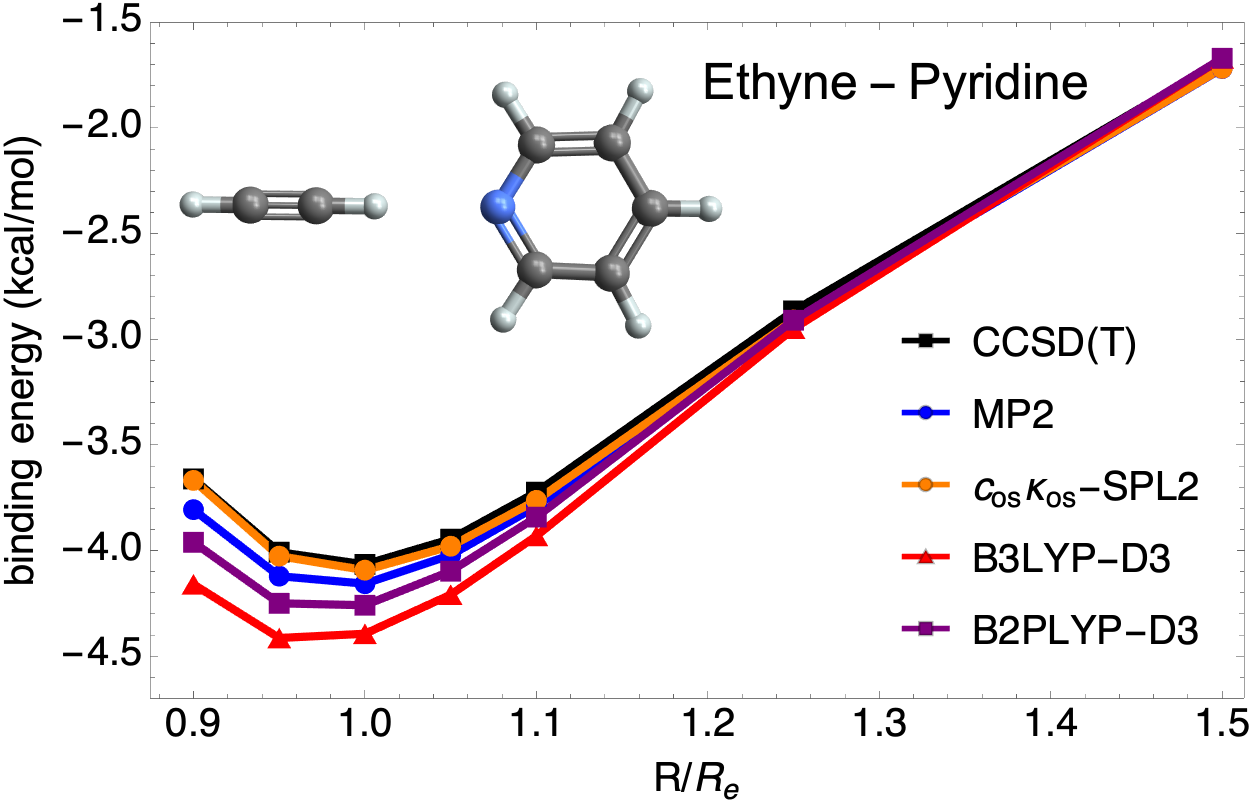}
\end{subfigure}
\begin{subfigure}{.49\textwidth}
    \centering
    \includegraphics[width=.95\linewidth]{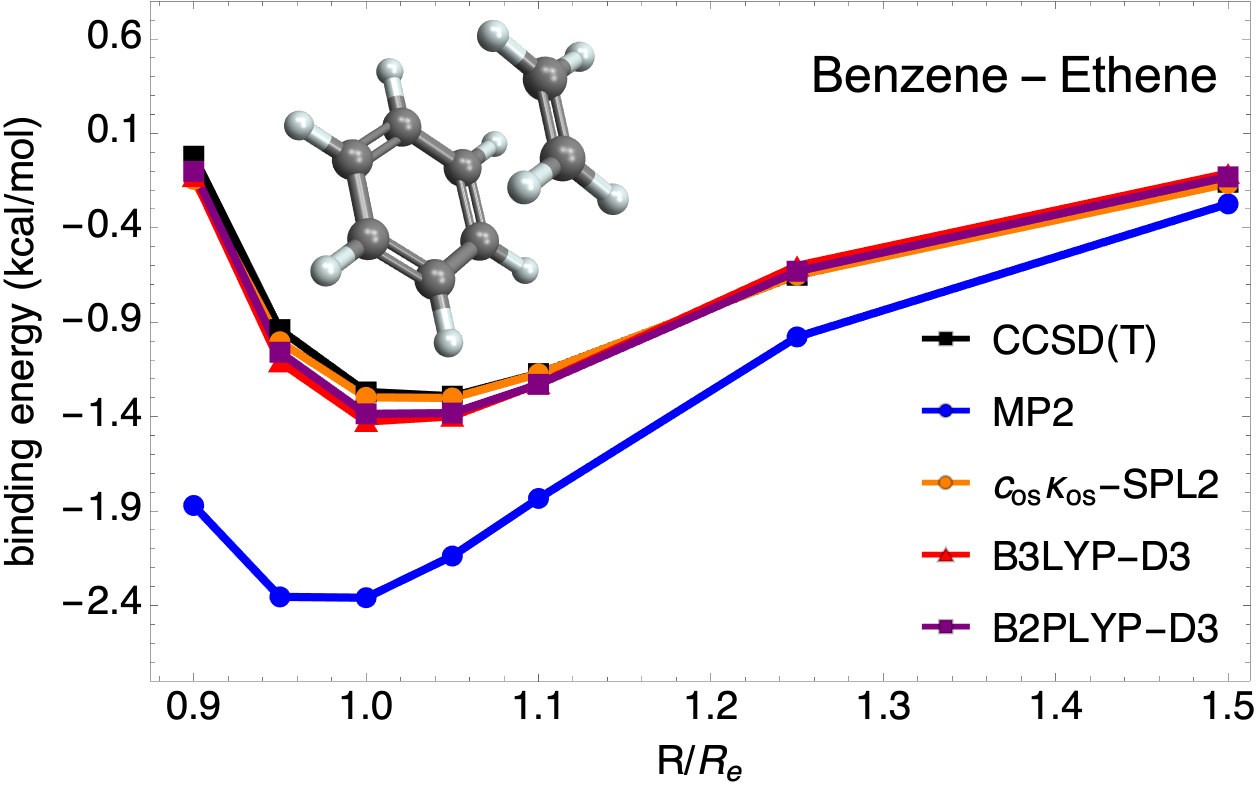}
\end{subfigure}
\begin{subfigure}{.49\textwidth}
    \centering
    \includegraphics[width=.95\linewidth]{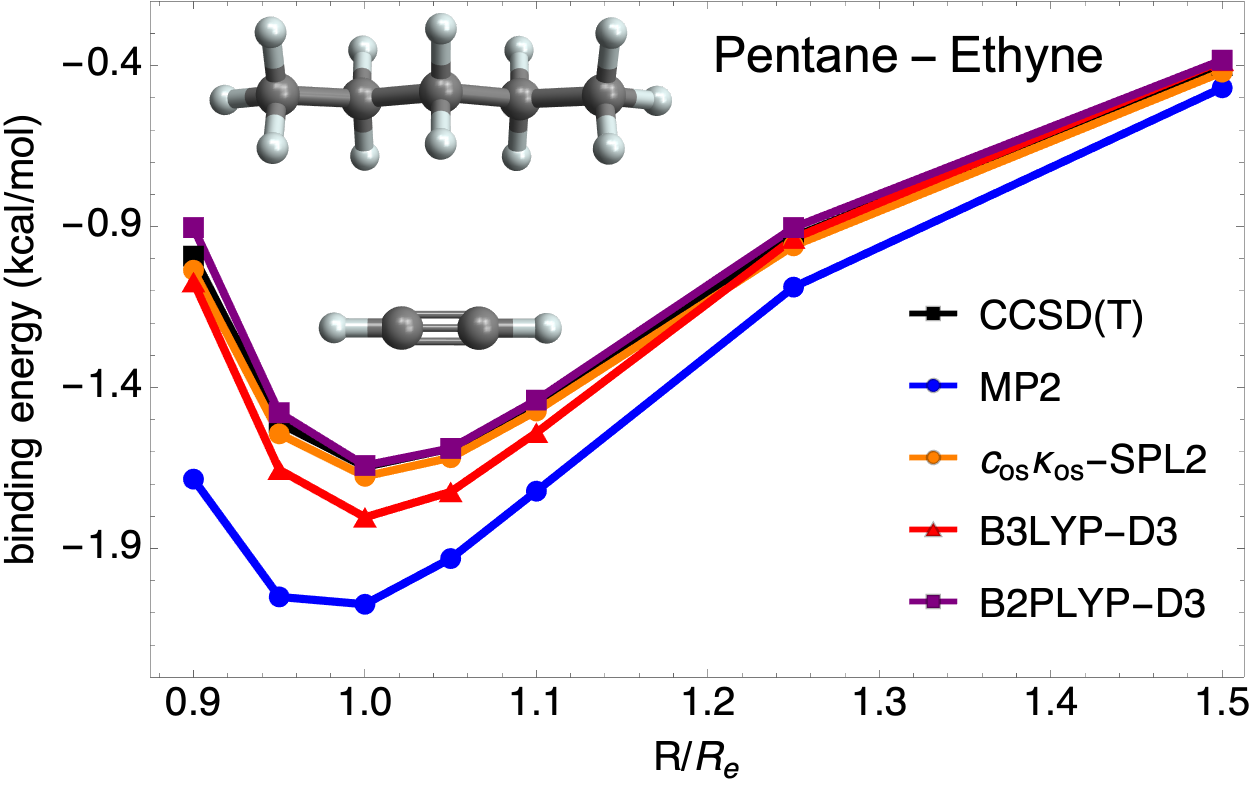}
\end{subfigure}
\caption{The dissociation curves of a dimer containing hydrogen bonds (top left), containing mixed interactions (top right), and two dimers containing dispersion interactions (bottom) for conventional wavefunction methods (CCSD(T) and MP2), dispersion corrected DFT functionals (B3LYP-D3 and B2PLYP-D3) and our new $c_{\rm os}\kappa_{\rm os}$-SPL2 functional. The dissociation curves of all the functionals can be found in the SI. (Figure~S5}
\label{fig:dis_curves}
\end{figure*}

Continuing with the S66x8 dataset, Fig.~\ref{fig:dis_curves} shows dissociation curves for four S66 complexes: AcOH - Uracil (representing hydrogen bonds), Ethyne - Pyridine (mixed interactions), and Benzene - Ethene I\& Pentane - Ethyne (both representing dispersion interactions).
The dissociation curves of $c_{\rm os}\kappa_{\rm os}$-SPL2 align closely with those of the CCSD(T) reference.
While B2PLYP-D3 also provides accurate curves,
it is outperformed by $c_{\rm os}\kappa_{\rm os}$-SPL2 in the case of the mixed Ethyne - Pyridine complex.
The $c_{\rm os}\kappa_{\rm os}$-SPL2 method effectively corrects the MP2 curves, both when MP2 under-binds (AcOH - Uracil),
and when it over-binds (remaining complexes).
In addition to errors in interaction energies,
MP2 also displays significant errors in the equilibrium distances between dispersion-bonded fragments
\cite{VucBur-JCPL-20,Vuc-JPCA-22} (position of the minimum in the dissociation curves), and this is also fixed by $c_{\rm os}\kappa_{\rm os}$-SPL2 (lower panels of Fig.~\ref{fig:dis_curves}).

\begin{table}[t]
    \centering
    \begin{tabular}{c||c|ccc}
 \text{Method} & \text{MAE} & \text{Hbonds} & \text{HGB} & \text{rest} \\\hline\hline
 \text{MP2} & 0.34 & 0.26 & 0.27 & 0.48 \\
 \text{SPL2} & 0.16 & 0.17 & 0.2 & 0.1 \\
 $\kappa -\text{F1}$ & 0.17 & 0.26 & 0.16 & 0.1 \\
 \text{B2PLYP-D3} & 0.18 & 0.25 & 0.13 & 0.16 \\\hline
 $c_{\text{\rm os}}\kappa _{\text{\rm os}}-\text{MP2}$ & 0.22 & 0.24 & 0.23 & 0.2 \\
 $c_{\text{\rm os}}\kappa _{\text{\rm os}}\text{-SPL2}$ & \textbf{0.1} & \textbf{0.1} & 0.12 & \textbf{0.09} \\
 $c_{\text{\rm os}}-\text{SPL2}$ & 0.12 & 0.15 & 0.11 & 0.1 \\
 \text{B3LYP-D3} & 0.22 & 0.46 & \textbf{0.1} & 0.13 \\
  \text{M06-2X} & 0.27 & 0.24 & 0.40 & 0.17 \\
  \end{tabular}
    \caption{The total MAE's of 31 complexes from the X40 dataset of 9 functionals split as well as the MAE's of the separate hydrogen bonding, halogen bonding, and the remaining NCIs. As in Ref. \onlinecite{MarHea-MP-17}, we have removed the Iodine complexes as they require relativistic corrections. This leaves the set with 31 different halogen bonded complexes, which have been also studied in the context of double hybrids\cite{MarHea-JCP-18,ForVis-JCC-20} and the $\kappa$-regularizers\cite{LoiBerLeeHea-JCTC-21,SheLoiRetLeeHea-JPCL-21}
    \label{tab:X40}.   Best result for each column is highlighted in boldface.}
\end{table}

 {\it \bf NCIs in Complexes Involving Halogens, Chalcogens, and Pnictogens.-}
We extend our analysis to NCI between molecules with halogen, chalcogen, and pnictogen atoms.
Simulations of halogen bonds pose significant challenges to dispersion-corrected DFT, with difficulties often ascribed to the {delocalization error} inherent 
to density-functional approximations, \cite{KimSonSimBur-JPCL-19, OteJohDil-JCTC-14, SonVucSimBur-JPCL-21, SonVucKimYuSimBur-Nat-23, MehFelWhiGoe-JCTC-21}
which cannot be fixed by empirical dispersion corrections such as D3 or D4. \cite{KimSonSimBur-JPCL-19, OteJohDil-JCTC-14, SonVucSimBur-JPCL-21, SonVucKimYuSimBur-Nat-23, MehFelWhiGoe-JCTC-21}
However, accurately capturing halogen bonds is crucial, given their significant role in numerous biomolecular (protein–ligand) complexes and diverse crystalline materials.\cite{RezRilHob-JCTC-12}
We test our methods against the X40 dataset from Rezac et al.,\cite{RezRilHob-JCTC-12} which includes a broad range of NCIs, spanning from hydrogen to pure halogen bonds in molecules containing halogens.
Table~\ref{tab:X40} provides a summary of the results for the X40 dataset, including the MAE for the overall X40 and for the hydrogen-bonded, halogen-bonded, and remaining complexes in the set.
Once again, $c_{\rm os}\kappa_{\rm os}$-SPL2 is the most accurate method, outperforming MP2 by factors ranging from 2 to 5 depending on the X40 category.
In comparison with B3LYP-D3, $c_{\rm os}\kappa_{\rm os}$-SPL2 exhibits comparable accuracy for halogen bonds and the ``remaining'' complexes, 
but surpasses B3LYP-D3 by a factor of $4.6$ for hydrogen bonds. 
We have added M06-2X~\cite{ZhaTru-TCA-08} to Table~\ref{tab:X40}, 
a functional that was assessed reliable for halogen-bonded complexes.\cite{OteJohDil-JCTC-14,AzeRamHamBic-JCC-21}
Our results demonstrate that $c_{\rm os}\kappa_{\rm os}$-SPL2 significantly outperforms M06-2X, both for the entire X40 dataset and for its  subsets.

\begin{figure}[t]
\centering
\includegraphics[width=.95\linewidth]{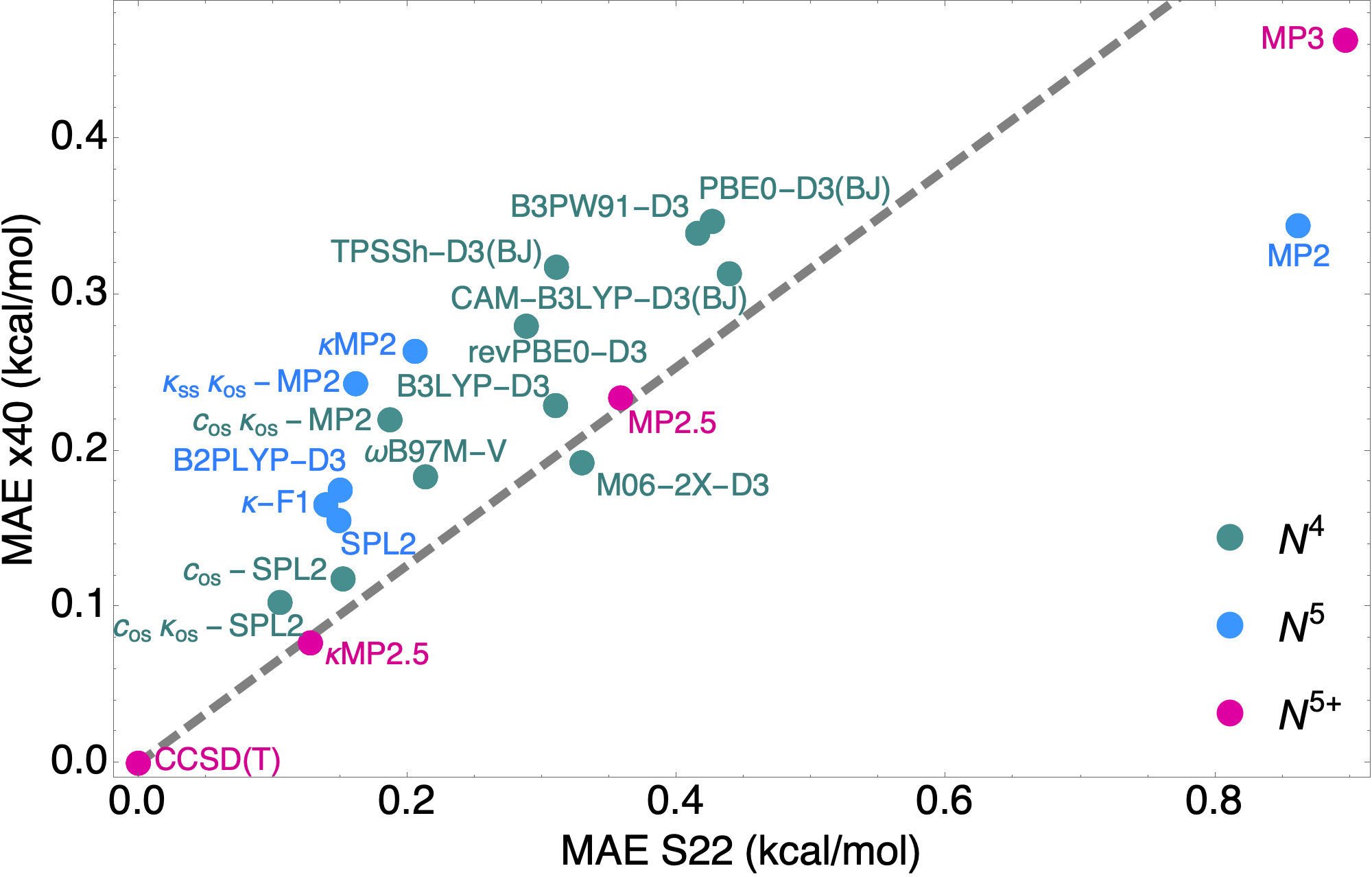}
\caption{Correlation plot for the X40 and S22 MAE for methods with 
$N^4$, $N^5$, $N^{5+}$ scalings. 
The results for all the hybrid functionals were taken from Ref.~\onlinecite{MarHea-MP-17}, the results for B2PLYP-D3 for S22 were taken from Ref.~\onlinecite{GoeHanBauEhrNajGri-RSC-2017} and for x40 were calculated using Turbomole (see computation details for more information) and the results for MP2.5, $\kappa$MP2.5 and MP3 were taken from Ref.~\onlinecite{LoiBerLeeHea-JCTC-21}.
}
\label{fig:x40s22}
\end{figure}

Before delving into more intricate halogen-bonded complexes, we remark that $c_{\rm os}\kappa_{\rm os}$-SPL2 displays remarkably low errors for both S22 and X40. The two set include NCIs of very
different nature, and yet, $c_{\rm os}\kappa_{\rm os}$-SPL2 attains MAEs of $0.11$ and $0.1$ kcal/mol, respectively (approaching the uncertainties in the 
CCSD(T) reference itself~\cite{RezDubJurHob-PCCP-15}).  
This level of accuracy, unequaled among methods of similar or even higher computational cost, is highlighted in Fig.\ref{fig:x40s22}. From this 
figure we can see that while 
$\kappa$-MP2.5 \cite{LoiBerLeeHea-JCTC-21} comes close to the performance of $c_{\rm os}\kappa_{\rm os}$-SPL2, it also comes
at a higher computational cost ($N^6$ scaling).

\begin{table}[]
    \centering
    \begin{tabular}{c||c|ccc}
 \text{Method} & \text{MAE} & \text{aHGB} & \text{CHB} & \text{PNB} \\\hline\hline
 \text{MP2} & 0.86 & 2.22 & 0.63 & \textbf{0.17} \\
 \text{SPL2} & 0.48 & 1.23 & \textbf{0.2} & 0.6 \\
 $\kappa -\text{F1}$ & 0.84 & 2.27 & 0.58 & 0.2 \\
 \text{B2PLYP-D3} & 0.72 & 2.02 & 0.23 & 0.89 \\\hline
 $c_{\text{\rm os}}\kappa _{\text{\rm os}}-\text{MP2}$ & 0.93 & 2.6 & 0.59 & 0.26 \\
 $c_{\text{\rm os}}\kappa _{\text{\rm os}}-\text{SPL2}$ & \textbf{0.45} & \textbf{0.73} & 0.4 & 0.31 \\
 $c_{\text{\rm os}}-\text{SPL2}$ & 0.9 & 1.53 & 0.64 & 1.03 \\
 \text{B3LYP-D3} & 1.36 & 3.50 & 0.98 & 0.36 \\
 \text{M06-2X} & 0.99 & 1.80 & 0.59 & 1.36\\
  \end{tabular}
    \caption{The total MAE's of the B30/Bauza30 dataset of 9 functionals split as well as the MAE's of the separate halogen bonding (containing anions as donors), chalcogen bonding and pnictogen bonding complexes. Best result for each column is highlighted in boldface.}
    \label{tab:B30}
\end{table}

The B30 dataset, comprising halogen, chalcogen, and pnictogen bonds\cite{BauAlkFroElg-JCTC-13}, has proven to be an even more difficult challenge for DFT than X40~\cite{MarHea-JCP-18,MarHea-MP-17},
being classified as ``difficult NCIs'' dataset.
In Table \ref{tab:B30}, the MAE of the full B30 set and the three separate subsets are displayed, and 
we can see that $c_{\rm os}\kappa_{\rm os}$-SPL2 outperforms other methods for anionic halogen bonds (aHGB), with a MAE of 0.73 kcal/mol, significantly surpassing all competing methods. 
In the case of chalcogen bonds, however, $c_{\rm os}\kappa_{\rm os}$-SPL2 is surpassed by SPL2, and by MP2 for pnictogen bonds. 
Nevertheless, for the full B30 set, $c_{\rm os}\kappa_{\rm os}$-SPL2 is still a top performer. 

\begin{figure*}[t]
    \centering
\includegraphics[width=.99\linewidth]{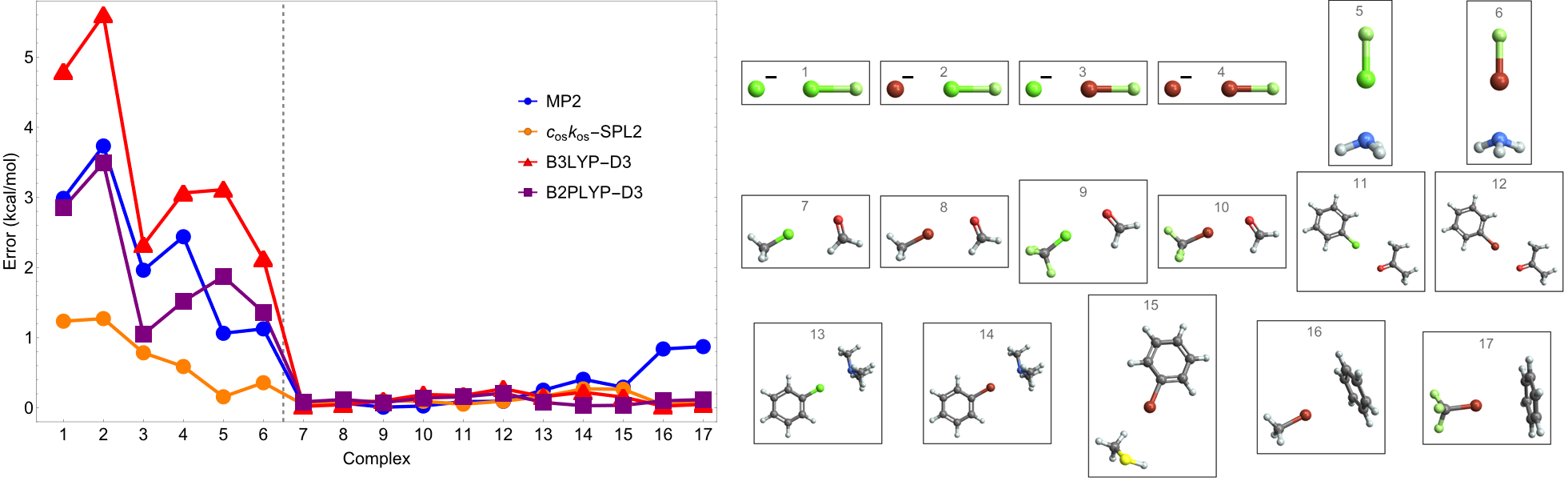}
\caption{The errors between the CCSD(T) reference data~\cite{OteJohDil-JCTC-14,RezRilHob-JCTC-12} and MP2, $c_{\rm os},\kappa_{\rm os}$-SPL2, B3LYP-D3 and B2PLYP-D3 respectively for the halogen bonded complexes of the B30~\cite{OteJohDil-JCTC-14,BauAlkFroElg-JCTC-13} and X40~\cite{RezRilHob-JCTC-12} datasets. The underlying complexes are shown on the right with the first four containing negatively charged halogen atoms. The curves for $c_{\rm os}\kappa_{\rm os}$-MP2 and SPL2 can be found in the SI (left panel of Figure~S6)}
\label{fig:hal_bonds}
\end{figure*}

\begin{figure}[t]
\centering
\includegraphics[width=.95\linewidth]{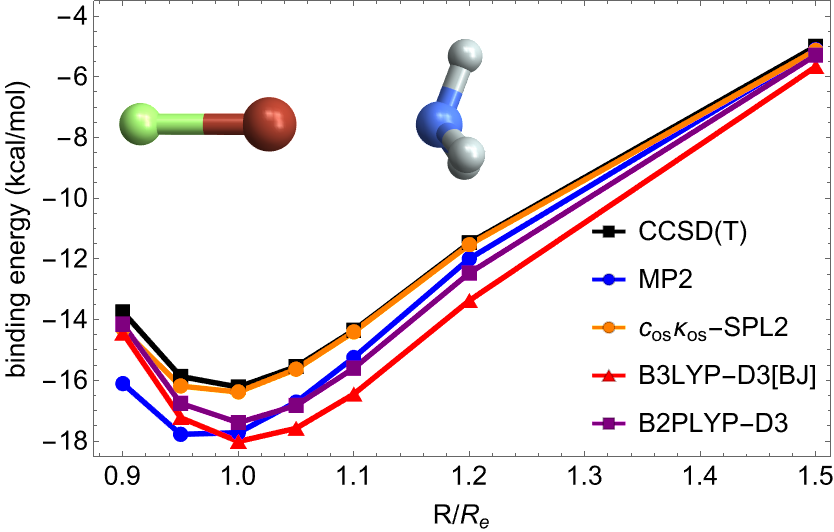}
\caption{The dissociation curve of \ce{NH3}-\ce{FBr} (complex 6) for CCSD(T), MP2, $c_{\rm os}\kappa_{\rm os}$-SPL2, B3LYP-D3[BJ] and B2PLYP-D3. Another dissociation curve of the B30 set can be found in the SI. (right panel of Figure~S6}
\label{fig:b30dis}
\end{figure}

The performance of $c_{\rm os}\kappa_{\rm os}$-SPL2 and the discrepancy in the accuracy of other methods for six B30 and eleven X40 halogen bonds are showcased in Fig.~\ref{fig:hal_bonds}. 
In B30, an electron donor in a halogen bond is an anion, 
whereas in X40 it is typically a lone electron pair on oxygen or nitrogen
within a neutral species. 
From Fig.~\ref{fig:hal_bonds}, we observe that both DFT-D3 and MP2 methods exhibit substantial errors for the 
more ``extreme'' B30 halogen-bond interactions. Yet, these methods demonstrate excellent performance for the remaining complexes, barring the MP2's performance for the last two, dominated by halogen-$\pi$ interactions.
For the B30 complexes in Fig.~\ref{fig:hal_bonds}, $c_{\rm os}\kappa_{\rm os}$-SPL2 is by far the most accurate, 
and its accuracy does not deteriorate once we move to the X40 complexes, 
suggesting that unlike other methods $c_{\rm os}\kappa_{\rm os}$-SPL2 
can accurately predict  halogen interactions of different nature. 
The good performance of DFT-D3 for 
X40 halogen bonds and very bad for B30 
is a consequence of large density-driven errors associated with the anions and charge-transfer prevalent in B30 halogen-bonded complexes. 
These errors have been extensively dissected in several studies.\cite{KimSonSimBur-JPCL-19, SimSonVucBur-JACS-22,SonVucSimBur-JCTC-22,SonVucSimBur-JPCL-21}

Furthermore, in Fig.~\ref{fig:b30dis}, we show the dissociation curves of \ce{NH3}-\ce{FBr} (complex 6 
in Fig.~\ref{fig:hal_bonds}). 
Here, $c_{\rm os}\kappa_{\rm os}$-SPL2 again outperforms the other functionals and closely matches the CCSD(T) reference. Even though B3LYP-D3 and B2PLYP-D3 at small bond distances and MP2 at larger ones display a good accuracy, $c_{\rm os}\kappa_{\rm os}$-SPL2 is the only method accurate around equilibrium. 
Similar conclusions can be drawn from the dissociation curve of another anionic halogen-bonded complex (see right panel of Fig. S6 in the SI). 

 {\it \bf Conclusion and Outlook.-} In summary, $c_{\rm os}\kappa_{\rm os}$-SPL2 offers an excellent performance for NCIs of various types. Building upon its SPL2 predecessor, $c_{\rm os}\kappa_{\rm os}$-SPL2 offers substantial improvements over both MP2 and dispersion-corrected (double) hybrid methods. Simultaneously, $c_{\rm os}\kappa_{\rm os}$-SPL2 achieves even greater accuracy than SPL2, while costing less ($N^4$ vs. $N^5$ scaling). Further, we demonstrate the robustness and versatility of the SPL2 correction, showing its capability to significantly correct not only ``bare'' MP2, for which it was initially designed, but also various MP2 modifications. This flexibility is maintained while retaining, or even improving upon, the original SPL2 accuracy. 
 While we used a large aug-cc-pVQZ basis set in this study~\cite{FabGorSeiDel-JCTC-16}, encouraged by SPL2's ability to absorb 
 errors of both MP2 and its alterations, 
 in the spirit of Refs.~\onlinecite{GinParFerAssSavTou-JCP-18,TraGinTou-JCP-22},
 we will also apply SPL2 to correct MP2 within smaller basis sets (e.g., that from Ref.~\onlinecite{MehMar-JPCA-22}).
 Finally, we will also use machine learning to create SPL2-like interpolations of the MP AC curve between its small and large $\lambda$ limits. This approach could offer even higher accuracy for NCIs. 

 \section{Computational Details}
For all calculations, PYSCF\cite{Sun-WIRCMS-18} is used to calculate the exchange energy, the HF density, and the MP2 correlation energy, which have been used for MP2, and SPL2 I\& F1 forms. To calculate the $\kappa$-MP2 correlation energies we used a python, PYSCF-like code with {\em numba} parallelization. This code and the full code for running MP AC calculations can be found in Ref.~\onlinecite{https://doi.org/10.5281/zenodo.8112015}. 
Unless otherwise specified, we have used the aug-cc-pVQZ\cite{Dun-JCP-89}.
The exceptions are the bromide complexes of X40\cite{RezRilHob-JCTC-12} for which we run \ce{Br} with a aug-cc-pVQZ-pp basisset\cite{PetFigGolStoDol-JCP-03} and the corresponding pseudo-potential\cite{KozMar-JCTC-13} as done by Mardirossian et al.\cite{MarHea-MP-17}. 
A frozen core was also applied in these cases. The iodine complexes were removed from X40 due to the need for relativistic corrections.

The reference data of S22 and S66x8 for B3LYP-D3 and B2PLYP-D3 can be found in Ref. \onlinecite{GoeHanBauEhrNajGri-RSC-2017} (def2-QZVP) and \onlinecite{BraKesKozMar-PCCP-16} (haVQZ), respectively. The NCCE31 data for both functionals are taken from Ref. \onlinecite{LiaNee-JCTC-15} (aug-cc-pVTZ) for B3LYP-D3 and B2PLYP-D3. 
The B3LYP-D3 and M06-2X data for B30 are from Ref. \onlinecite{MarHea-MP-17} (aug-cc-pVTZ) and \onlinecite{LoiBerLeeHea-JCTC-21} (aug-cc-pVTZ), whereas the data for B2PLYP-D3 have been calculated using TURBOMOLE\cite{TURBOMOLE,Fur-WIR-14} with an aug-cc-pVQZ basis. 
For X40 the B3LYP-D3 are from Ref. \onlinecite{MarHea-MP-17} (aug-cc-pVTZ) and the B2PLYP-D3 data have been calculated from the aug-cc-pVTZ basis set in TURBOMOLE. T
The data from the dissociation curves for MP2 and CCSD(T) have been also calculated with TURBOMOLE and extrapolated to the CBS limit. 
The B2PLYP-D3 and B3LYP-D3 dissociation curves have been calculated using a standard aug-cc-pVQZ basis set in TURBOMOLE, whereas the $c_{\rm os}\kappa_{\rm os}$-SPL2 has been run within the same basis set in PYSCF.
The fitted parameters of all the functionals, other than the ones from Tab.~\ref{tab:funcs}, can be found in Table~S2 of the SI.

\section*{Acknowledgements}
SV acknowledges funding from the SNSF Starting Grant project (TMSGI2\_211246).
KJD, DPK, and PG-G acknowledge Financial support from the Netherlands Organisation for Scientific Research under Vici grant 724.017.001.
F.D.S. thanks the financial support from ICSC – Centro Nazionale di Ricerca in High Performance Computing, Big Data and Quantum Computing, funded by European Union – NextGenerationEU - PNRR.
We want to thank Arno Förster for calculating the aug-QZ6P B2PLYP-D3 data for X40 and for fruitful discussions regarding this work. 
We also want to thank Suhwan Song for providing his MP2 data for the B30 dataset  and Lucas de Azevedo Santos for insightful discussions. 

\section*{Associated Content}
The Supporting Information is available free of charge at
\begin{itemize}
    \item {Additional S66x8 Plots, parameters for the new methods, additional dissociation curves and results for halogen bonds}
\end{itemize}

\section*{Data Availability Statement}

The data that support the findings of this study are available within the main manuscript and its Supplemental Information. Additional raw data and the code can also be found on Zenodo in Ref.~\onlinecite{https://doi.org/10.5281/zenodo.8118099} and Ref.~\onlinecite{https://doi.org/10.5281/zenodo.8112015} respectively. 
\bibliographystyle{achemso}
\bibliography{bib_clean}

\end{document}


\author{Kimberly J. Daas}
\affiliation
{Department of Chemistry \& Pharmaceutical Sciences and Amsterdam Institute of Molecular and Life Sciences (AIMMS), Faculty of Science, Vrije Universiteit, De Boelelaan 1083, 1081HV Amsterdam, The Netherlands}
\author{Derk P. Kooi}
\affiliation
{Department of Chemistry \& Pharmaceutical Sciences and Amsterdam Institute of Molecular and Life Sciences (AIMMS), Faculty of Science, Vrije Universiteit, De Boelelaan 1083, 1081HV Amsterdam, The Netherlands}
\affiliation{Microsoft Research AI4 Science, Microsoft Research, Evert van de Beekstraat 354, 1118CZ Schiphol, The Netherlands}
\author{Nina C. Peters}
\affiliation
{Department of Chemistry \& Pharmaceutical Sciences and Amsterdam Institute of Molecular and Life Sciences (AIMMS), Faculty of Science, Vrije Universiteit, De Boelelaan 1083, 1081HV Amsterdam, The Netherlands}
\author{Eduardo Fabiano}
\affiliation{Institute for Microelectronics and Microsystems (CNR-IMM), Via Monteroni, Campus Unisalento, 73100 Lecce, Italy}
\author{Fabio Della Sala}
\affiliation{Institute for Microelectronics and Microsystems (CNR-IMM), Via Monteroni, Campus Unisalento, 73100 Lecce, Italy}
\author{Paola Gori-Giorgi}
\affiliation{Department of Chemistry \& Pharmaceutical Sciences and Amsterdam Institute of Molecular and Life Sciences (AIMMS), Faculty of Science, Vrije Universiteit, De Boelelaan 1083, 1081HV Amsterdam, The Netherlands}
\affiliation{Microsoft Research AI4 Science, Microsoft Research, Evert van de Beekstraat 354, 1118CZ Schiphol, The Netherlands}
\author{Stefan Vuckovic}
\affiliation{Department of Chemistry, Faculty of Science and Medicine, Université de Fribourg/Universität Freiburg, Chemin du Musée 9, CH-1700 Fribourg, Switzerland}
\title{Supplementary Information of: Regularized and scaled Opposite-spin Functionals in M{\o}ller-Plesset Adiabatic Connection: Higher Accuracy at a Lower Cost}
\maketitle

\section{Extra graphs and tables main text}

\begin{figure}
\centering
\includegraphics[width=.95\linewidth]{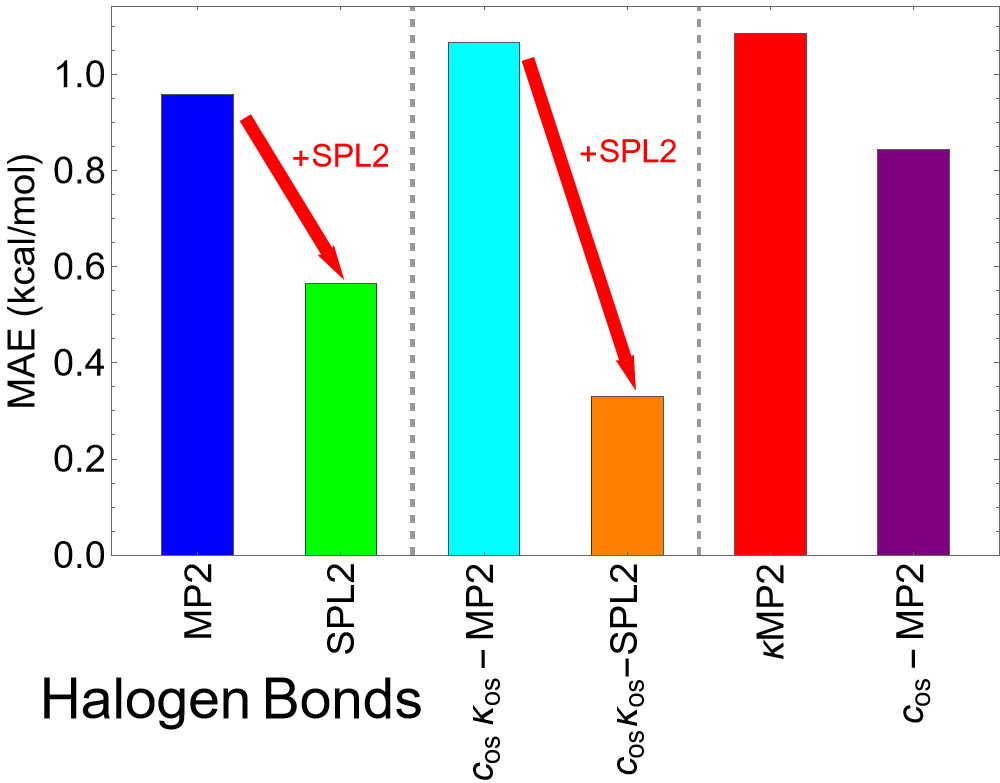}
\caption{The MAE (Mean Absolute Error) of MP2, SPL2, $c_{\rm os}\kappa_{\rm os}$-MP2 and $c_{\rm os}\kappa_{\rm os}$-SPL2, $\kappa$MP2 and $c_{\rm os}$-MP2 for the halogen bonded complexes of B30\cite{BauAlkFroElg-JCTC-13} and X40\cite{RezRilHob-JCTC-12} (see Fig.~8 for more detail on the complexes and Table~\ref{tab:funcs} for individual methods)}
\label{fig:halobondsMP2}
\end{figure}

\begin{table*}[]
\caption{The 20 different combinations of functionals that we have studied in this work. The bolded red functionals were also shown in Tab.I of the paper.}
\begin{tabular}{l||p{25mm}p{25mm}p{30mm}|p{26mm}p{26mm}}
            & Method & $\kappa-$Method & $\kappa_{\rm ss}\kappa_{\rm os}-$Method & $c_{\rm os}\kappa_{\rm os}-$Method & $c_{\rm os}-$Method\\ \hline\hline
MP2         & \br{$c_{\rm ss}=c_{\rm os}=1$}\newline \br{$\kappa_{\rm ss}=\kappa_{\rm os}=\infty$}    & $c_{\rm ss}=c_{\rm os}=1$\newline $\kappa_{\rm ss}=\kappa_{\rm os}=1.1$      &  $c_{\rm ss}=c_{\rm os}=1$\newline $\kappa_{\rm ss}=0.9,\kappa_{\rm os}=1.4$         &   \br{$c_{\rm ss}=0,c_{\rm os}=2.1$\newline $\kappa_{\rm ss}=0,\kappa_{\rm os}=0.9$}           & $c_{\rm ss}=0,c_{\rm os}=1.7$\newline $\kappa_{\rm ss}=\kappa_{\rm os}=\infty$       \\\hline 
SPL2        & \br{$c_{\rm ss}=c_{\rm os}=1$\newline $\kappa_{\rm ss}=\kappa_{\rm os}=\infty$}    &  $c_{\rm ss}=c_{\rm os}=1$\newline $\kappa_{\rm ss}=\kappa_{\rm os}=1.7$     &  $c_{\rm ss}=c_{\rm os}=1$\newline $\kappa_{\rm ss}=1.1,\kappa_{\rm os}=1.7$              &  \br{$c_{\rm ss}=0,c_{\rm os}=2.1$\newline $\kappa_{\rm ss}=0,\kappa_{\rm os}=1.3$}             &   \br{$c_{\rm ss}=0,c_{\rm os}=1.8$\newline $\kappa_{\rm ss}=\kappa_{\rm os}=\infty$}        \\\hline 
MPACF1          & $c_{\rm ss}=c_{\rm os}=1$\newline $\kappa_{\rm ss}=\kappa_{\rm os}=\infty$    &  $c_{\rm ss}=c_{\rm os}=1$\newline $\kappa_{\rm ss}=\kappa_{\rm os}=1.3$     &   $c_{\rm ss}=c_{\rm os}=1$\newline $\kappa_{\rm ss}=1,\kappa_{\rm os}=1.4$             &  $c_{\rm ss}=0,c_{\rm os}=2.3$\newline $\kappa_{\rm ss}=0,\kappa_{\rm os}=1.1$             &      $c_{\rm ss}=0,c_{\rm os}=2.2$\newline $\kappa_{\rm ss}=\kappa_{\rm os}=\infty$     \\\hline 
F1 &  $c_{\rm ss}=c_{\rm os}=1$\newline $\kappa_{\rm ss}=\kappa_{\rm os}=\infty$   & \br{$c_{\rm ss}=c_{\rm os}=1$\newline $\kappa_{\rm ss}=\kappa_{\rm os}=1.5$ }     &    $c_{\rm ss}=c_{\rm os}=1$\newline $\kappa_{\rm ss}=1.6,\kappa_{\rm os}=1.3$            & $c_{\rm ss}=0,c_{\rm os}=2.4$\newline $\kappa_{\rm ss}=0,\kappa_{\rm os}=1.0$              & $c_{\rm ss}=0,c_{\rm os}=2.0$\newline $\kappa_{\rm ss}=\kappa_{\rm os}=\infty$      
\end{tabular}
\label{tab:funcs}
\end{table*}

\begin{table*}[]
\caption{The 20 different combinations of functionals that we have studied in this work. The bolded red functionals have been studied in the paper.}
\begin{tabular}{l|l}
             Method & fitted parameters\\ \hline
\br{SPL2}& \br{$b_2=0.117$, $m_2 =10.68$, $\alpha=1.1472$, $\beta=-0.7397$}\\
$\kappa$-SPL2 & $b_2=-0.433$, $m_2 =5.775$, $\alpha=1.843$, $\beta=-1.750$\\
$\kappa_{\rm ss}\kappa_{\rm os}$-SPL2 & $b_2=-0.690$, $m_2 =3.831$, $\alpha=3.382$, $\beta=-4.026$  \\
\br{$c_{\rm os}\kappa_{\rm os}$-SPL2} &  \br{$b_2=0.287$, $m_2 =148.982$, $\alpha=1.674$, $\beta=-1.973$}\\
\br{$c_{\rm os}$-SPL2} & \br{$b_2=0.527$, $m_2 =58.850$, $\alpha=1.278$, $\beta=-1.059$} \\
MPACF1 & $d_1=0.294$, $d_2=0.934$ \\
$\kappa$-MPACF1 & $d_1=-0.3660$, $d_2=0.4677$\\
$\kappa_{\rm ss}\kappa_{\rm os}$-MPACF1 & $d_1=0.0001615$, $d_2=-0.0151$ \\
$c_{\rm os}\kappa_{\rm os}$-MPACF1 & $d_1=0.9965$, $d_2=0.6799$\\
$c_{\rm os}$-MPACF1 & $d_1=2.206$, $d_2=0.7068$ \\
F1 & $d_1=2.151$, $d_2 =0.413$, $\alpha=3.837$, $\beta=-6.620$ \\
\br{$\kappa$-F1} & \br{$d_1=1.147$, $d_2 =-0.6191$, $\alpha=2.279$, $\beta=-4.989$} \\
$\kappa_{\rm ss}\kappa_{\rm os}$-F1 & $d_1=0.398$, $d_2 =0.663$, $\alpha=2.715$, $\beta=-3.982$ \\
$c_{\rm os}\kappa_{\rm os}$-F1 & $d_1=1.380$, $d_2 =-0.5590$, $\alpha=2.902$, $\beta=-7.836$ \\
$c_{\rm os}$-F1 & $d_1=2.769$, $d_2 =-0.3665$, $\alpha=8.3970$, $\beta=-14.2015$\\
\end{tabular}
\label{tab:funcs_conts}
\end{table*}

\begin{figure}
\centering
\includegraphics[width=.95\linewidth]{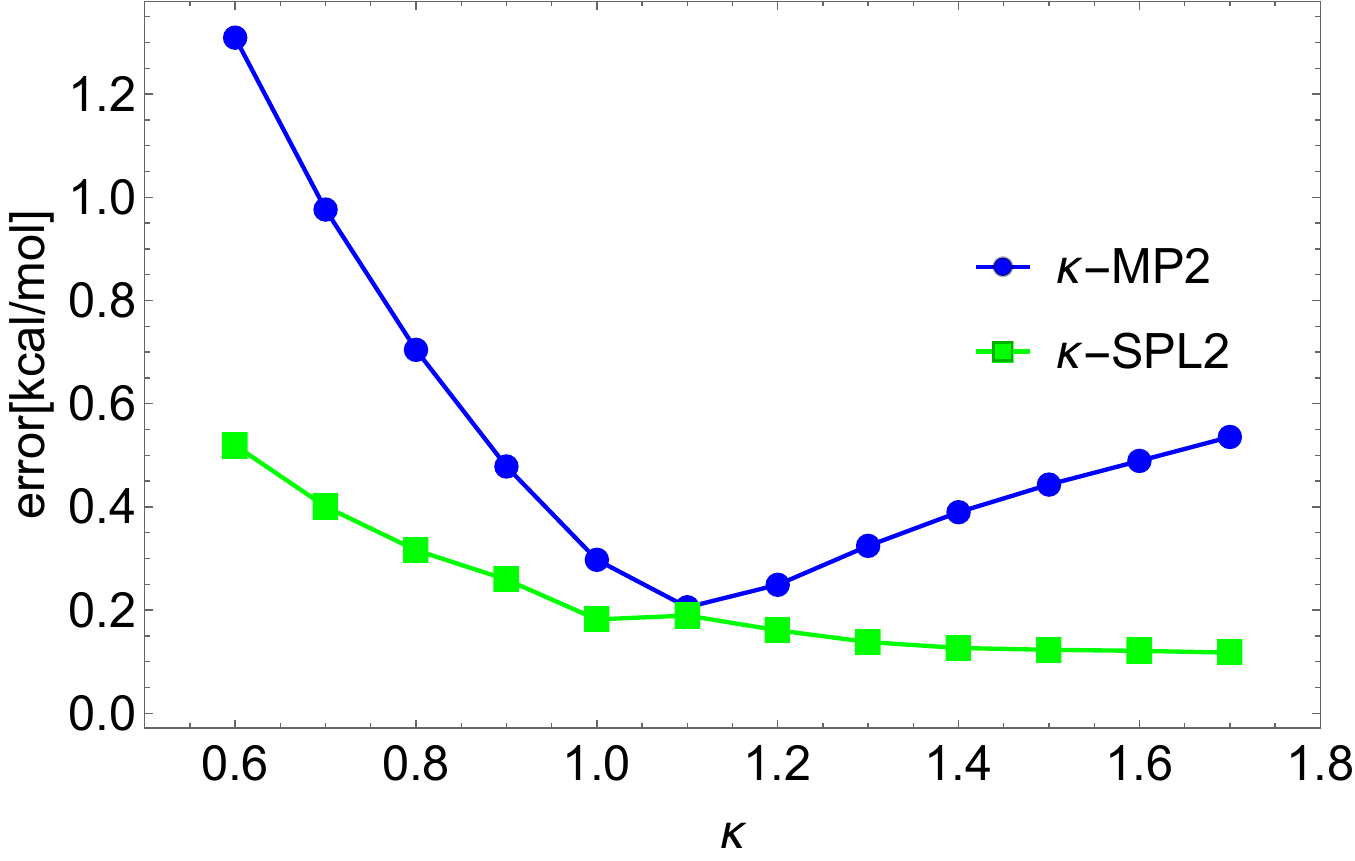}
\caption{The error between the interaction energy of $\kappa$-MP2 and $\kappa$-SPL2 and the CCSD(T) reference data for a range of $\kappa$'s between 0.6 and 1.7 of the S22 dataset.}
\label{fig:kappaS22}
\end{figure}

\begin{table}[]
    \centering
    \begin{tabular}{c||c|ccccc}
 \text{Method} & \text{MAE} & \text{HB6} & \text{CT7} & \text{DI6} & \text{WI7} & \text{PPS5} \\\hline\hline
 \text{MP2} & 0.5 & \textbf{0.42} & 0.86 & 0.46 & 0.04 & 0.79 \\
 \text{SPL2} & 0.25 & 0.54 & 0.41 & 0.18 & 0.04 & 0.08 \\
 $\kappa -\text{F1}$ & 0.32 & 0.73 & 0.52 & 0.18 & 0.03 & 0.11 \\
 \text{B2PLYP-D3} & 0.61 & 0.51 & 1.05 & 0.77 & \textbf{0.02} & 0.73 \\\hline
 $c_{\text{\rm os}}\kappa _{\text{\rm os}}-\text{MP2}$ & 0.4 & 0.77 & 0.67 & 0.26 & 0.06 & 0.24 \\
 $c_{\text{\rm os}}\kappa _{\text{\rm os}}-\text{SPL2}$ & 0.27 & 0.55 & 0.47 & 0.18 & 0.04 & 0.11 \\
 $c_{\text{\rm os}}-\text{SPL2}$ & \textbf{0.21} & 0.47 & \textbf{0.25} & \textbf{0.17} & 0.09 & \textbf{0.07} \\
 \text{B3LYP-D3} & 0.60 & 0.46 & 1.43 & 0.71 & 0.05 & 0.28 \\
    \end{tabular}
    \caption{The total MAE's of the NCCE31 dataset of 8 functionals split as well as the MAE's of the separate sets containing hydrogen bonds, charge transfer interaction, dipole interactions, weak interactions, and $\pi$-$\pi$-stacking interactions.}
    \label{tab:NCCE31}
\end{table}

\subsection{Additional S66x8 Plots}
\begin{figure*}
\begin{subfigure}{.49\textwidth}
    \centering
    \includegraphics[width=.95\linewidth]{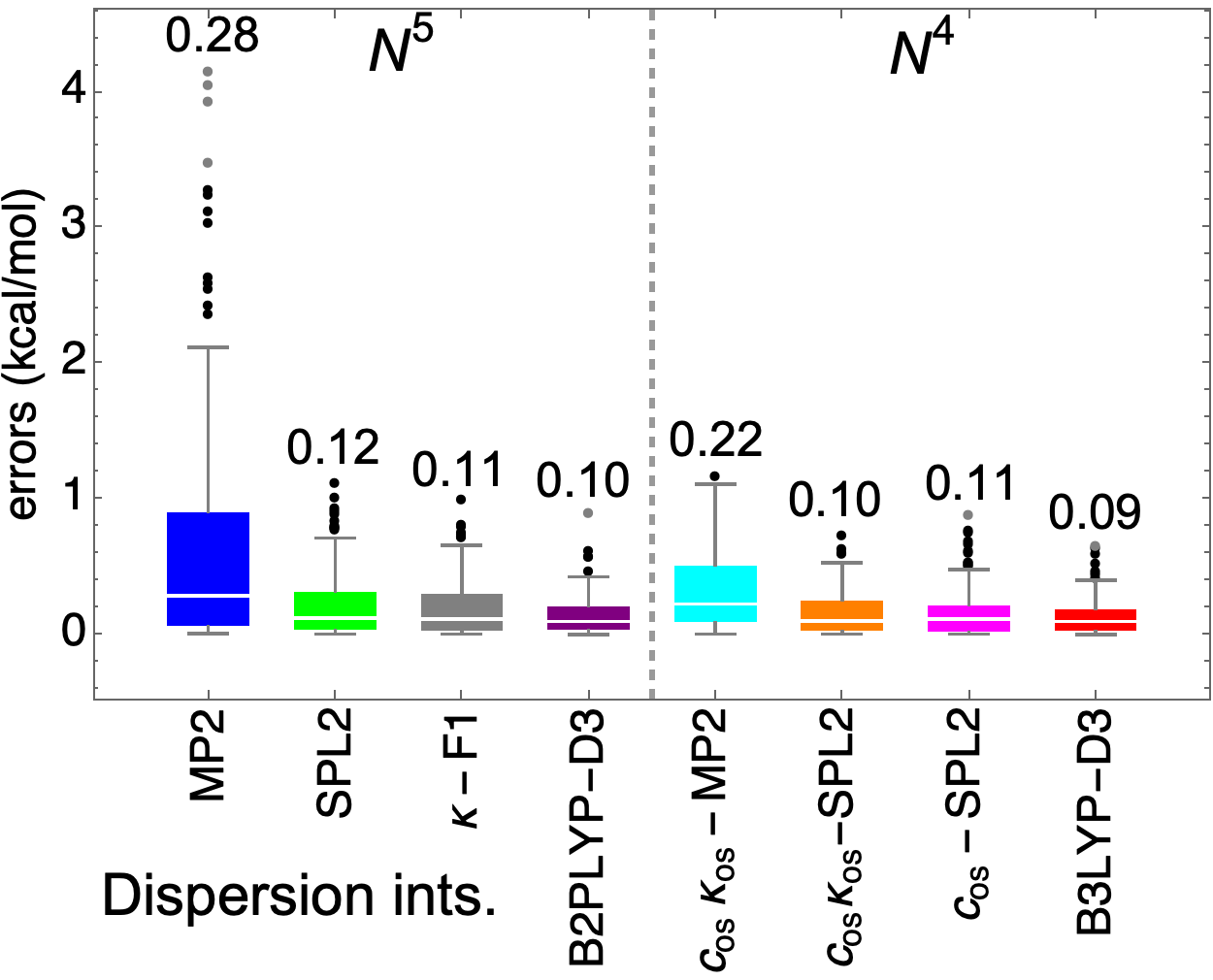}
\end{subfigure}
\begin{subfigure}{.49\textwidth}
    \centering
    \includegraphics[width=.95\linewidth]{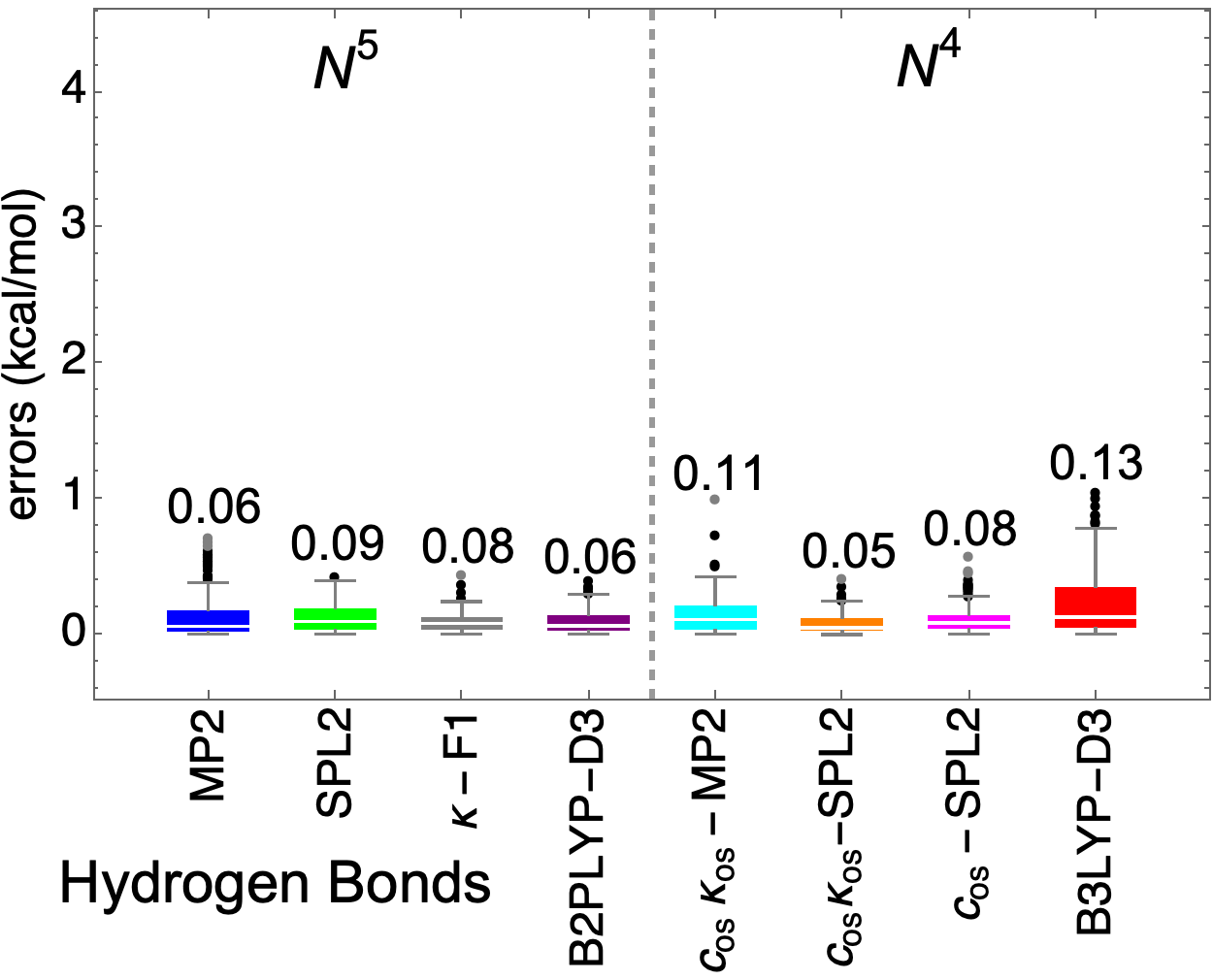}
\end{subfigure}
\begin{subfigure}{.49\textwidth}
    \centering
    \includegraphics[width=.95\linewidth]{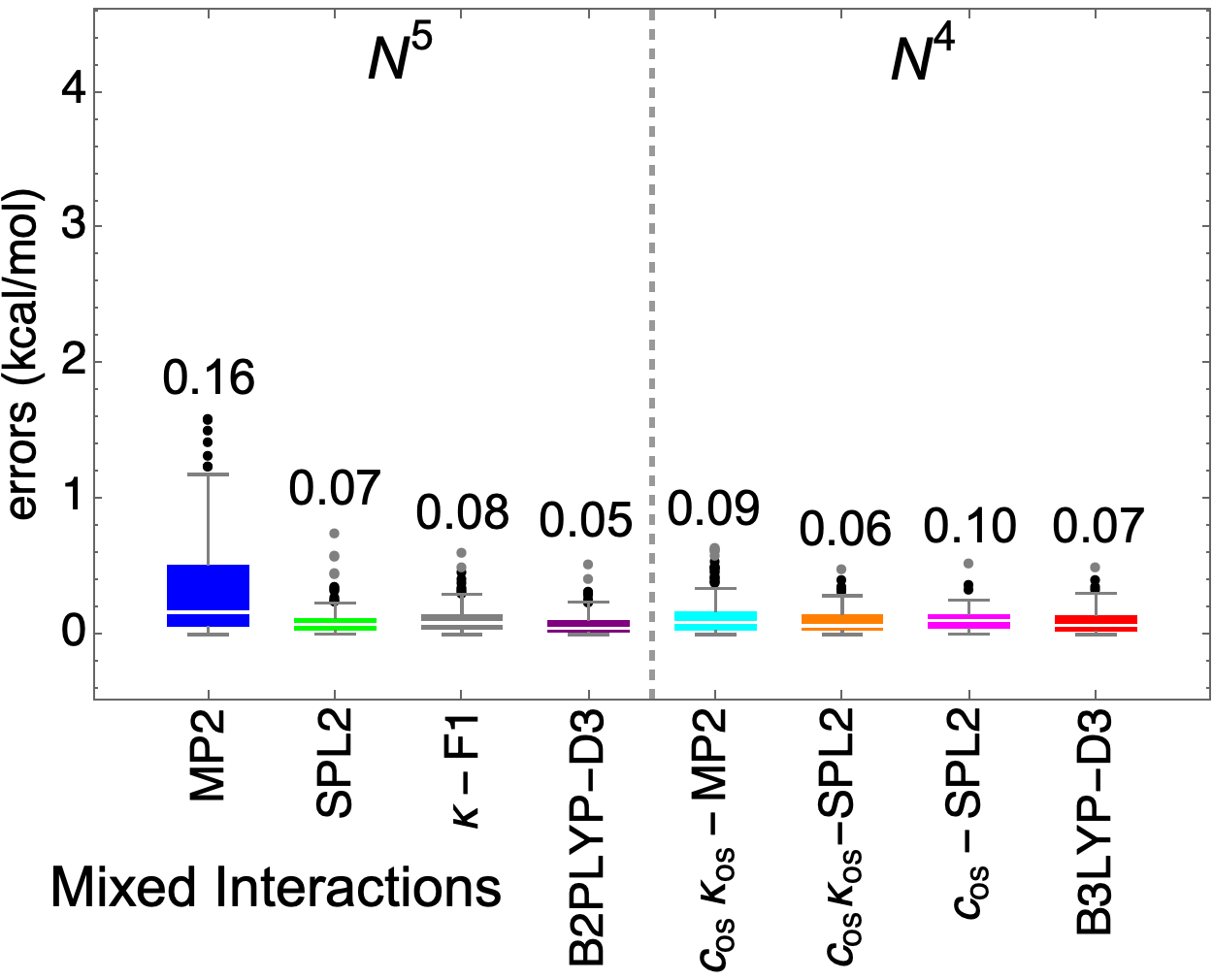}
\end{subfigure}
\begin{subfigure}{.49\textwidth}
    \centering
    \includegraphics[width=.95\linewidth]{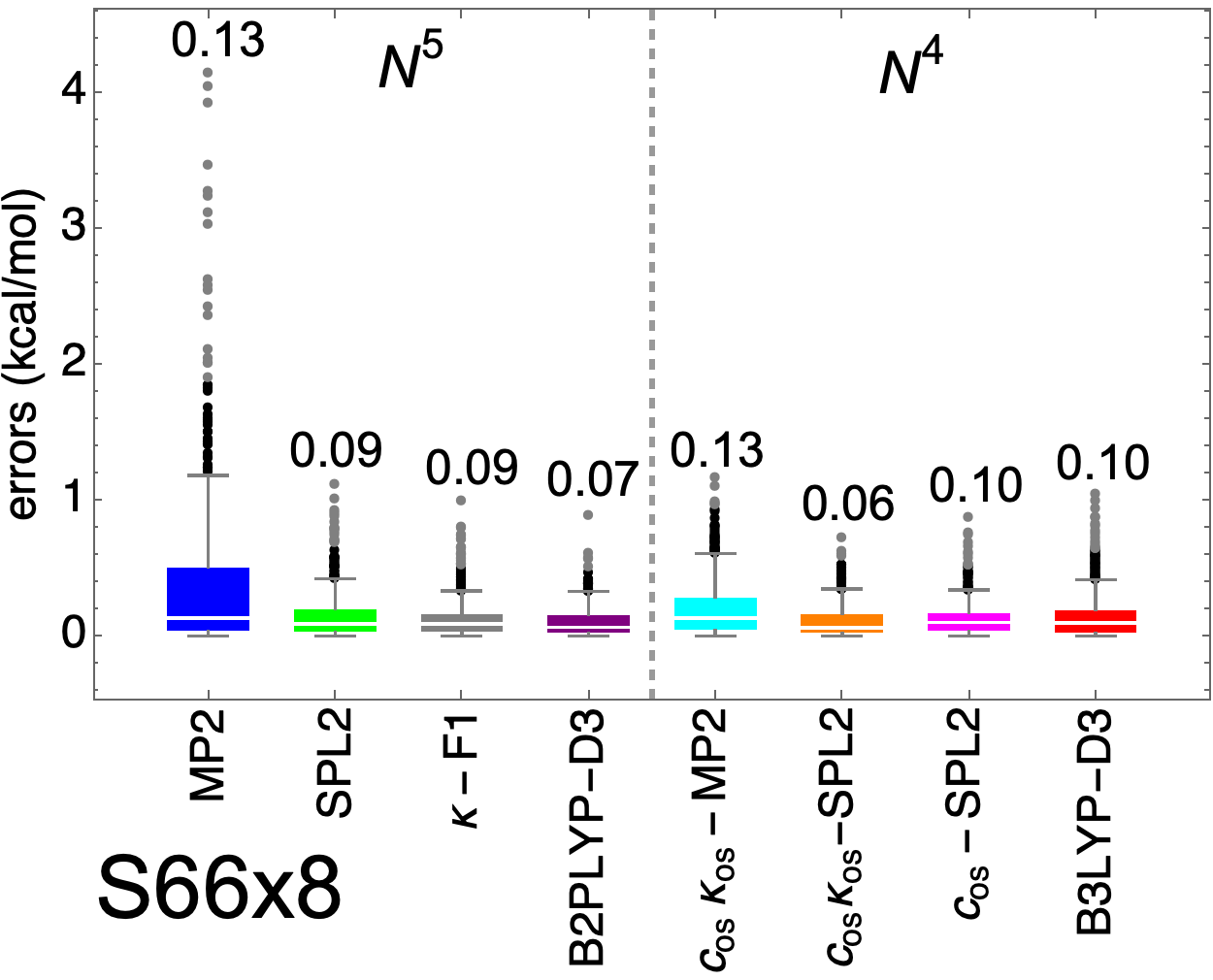}
\end{subfigure}
\caption{The box-plots containing the errors of all the 8 functionals for the S66x8 dataset split into the 3 different kind of interactions and the full dataset.}
\label{fig:all_s66}
\end{figure*}

\begin{figure*}
\begin{subfigure}{.49\textwidth}
    \centering
    \includegraphics[width=.95\linewidth]{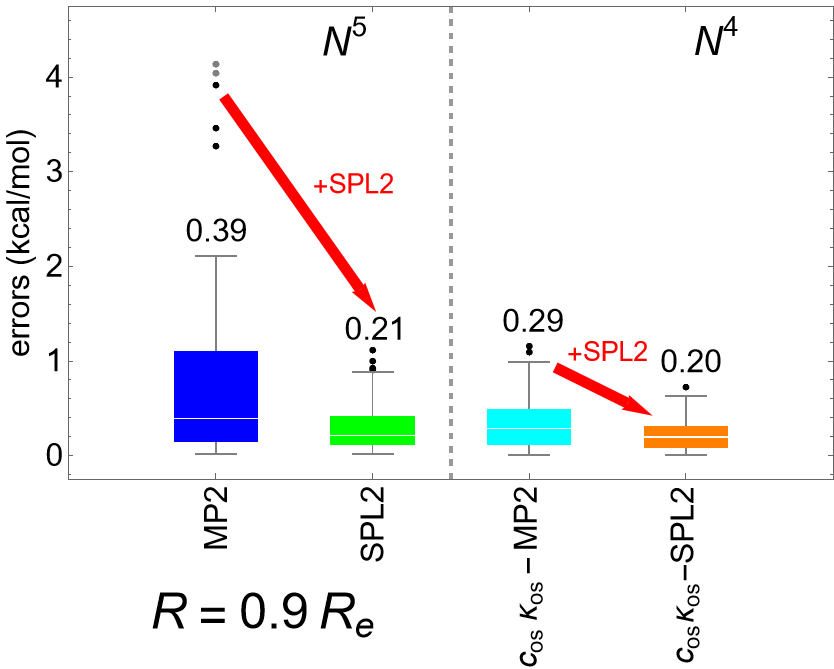}
\end{subfigure}
\begin{subfigure}{.49\textwidth}
    \centering
    \includegraphics[width=.95\linewidth]{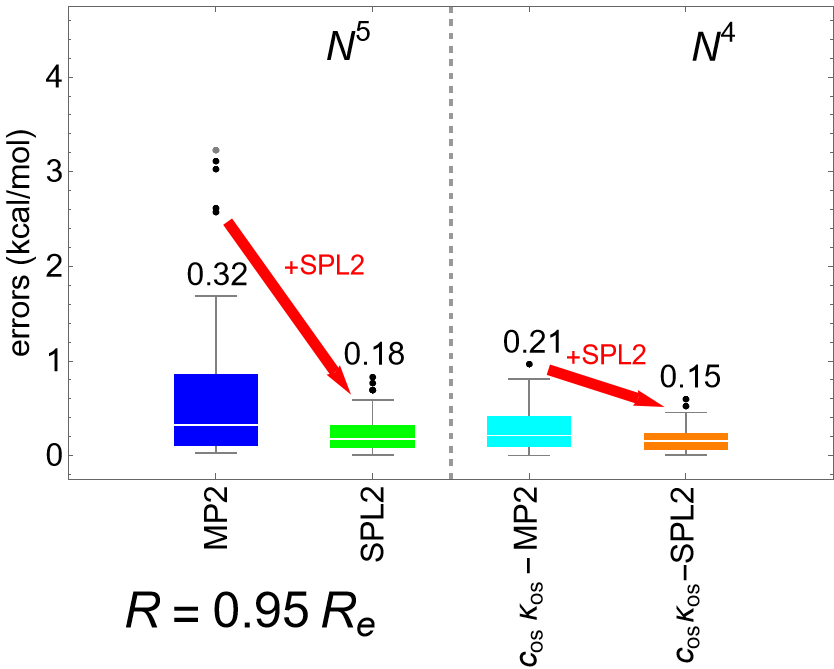}
\end{subfigure}
\begin{subfigure}{.49\textwidth}
    \centering
    \includegraphics[width=.95\linewidth]{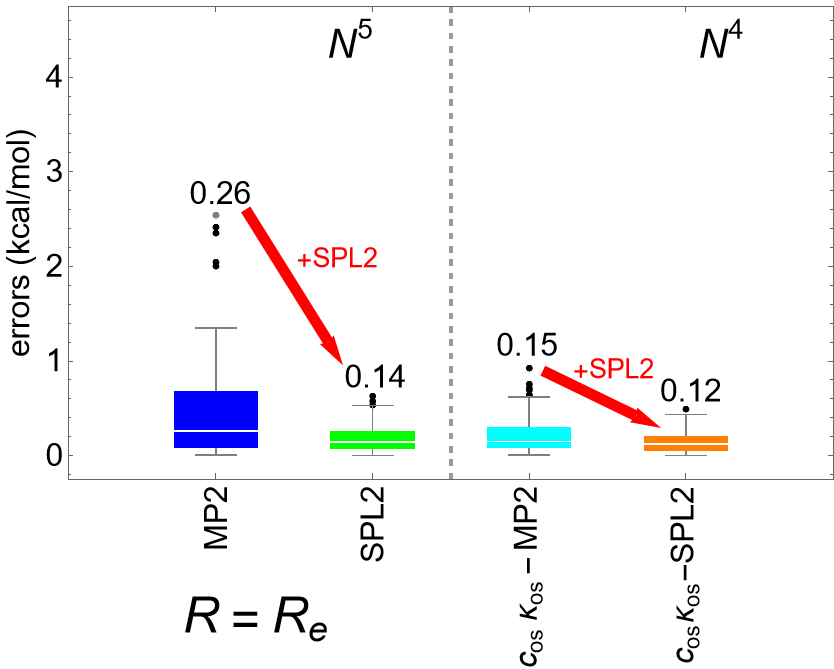}
\end{subfigure}
\begin{subfigure}{.49\textwidth}
    \centering
    \includegraphics[width=.95\linewidth]{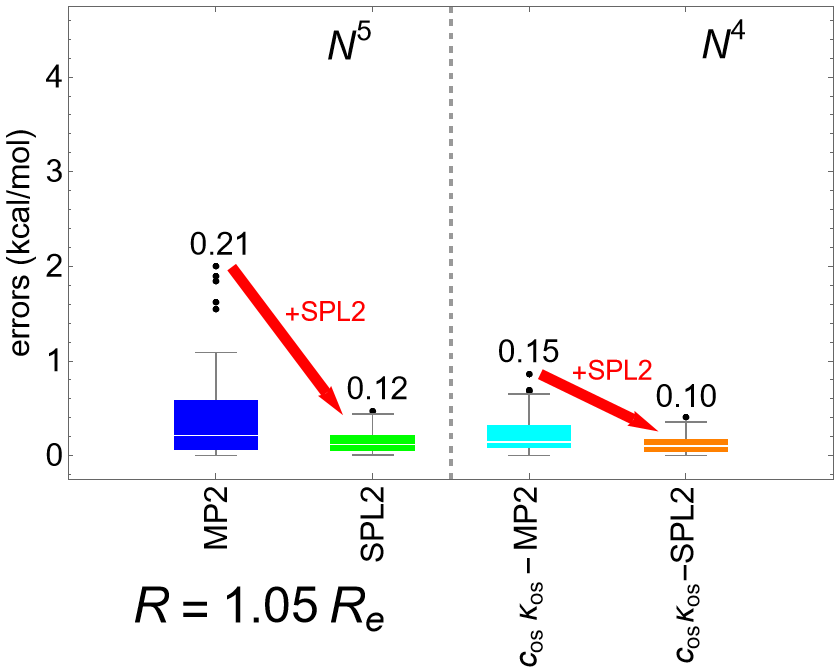}
\end{subfigure}
\caption{The box-plots containing the errors of MP2, SPL2 and $c_{\rm os}\kappa_{\rm os}$-MP2 and $c_{\rm os}\kappa_{\rm os}$-MP2 for the positions along the dissociation curve around the equilibrium distance, where it can be seen that the SPL2 corrections can fix the errors of MP2 forms even at small distances.}
\label{fig:Box_Pos}
\end{figure*}

\begin{figure*}
\begin{subfigure}{.49\textwidth}
    \centering
    \includegraphics[width=.95\linewidth]{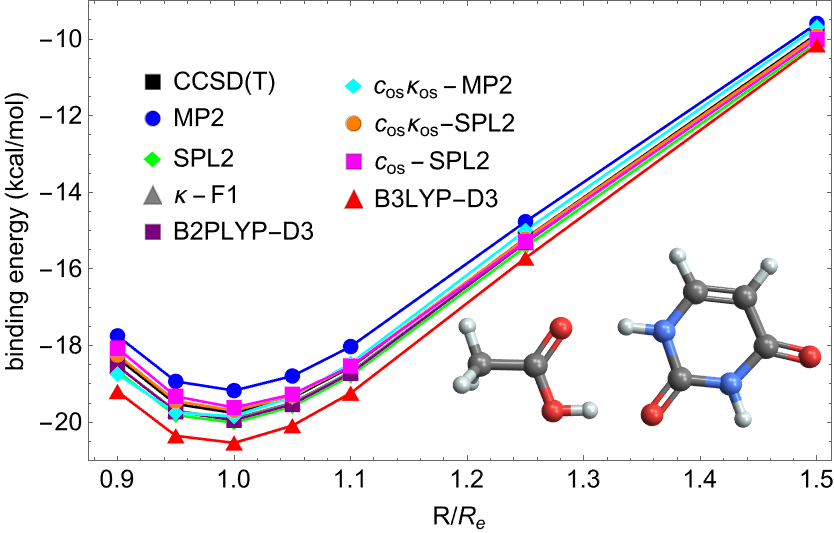}
\end{subfigure}
\begin{subfigure}{.49\textwidth}
    \centering
    \includegraphics[width=.95\linewidth]{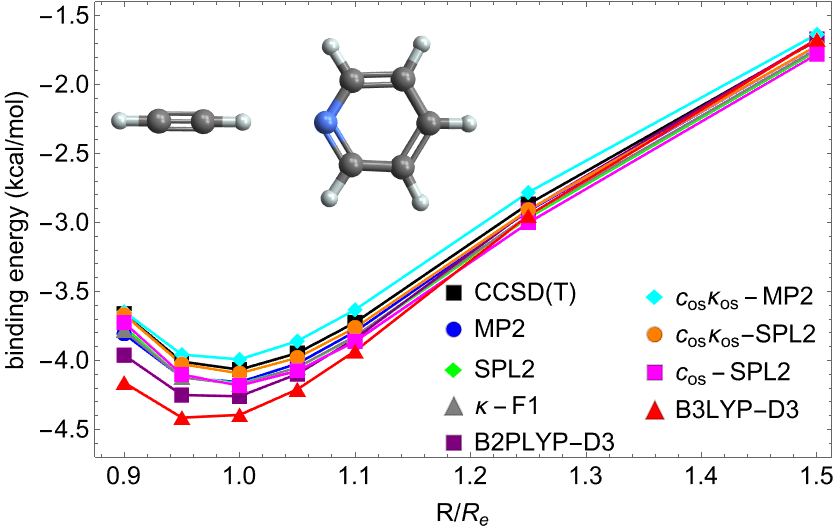}
\end{subfigure}
\begin{subfigure}{.49\textwidth}
    \centering
    \includegraphics[width=.95\linewidth]{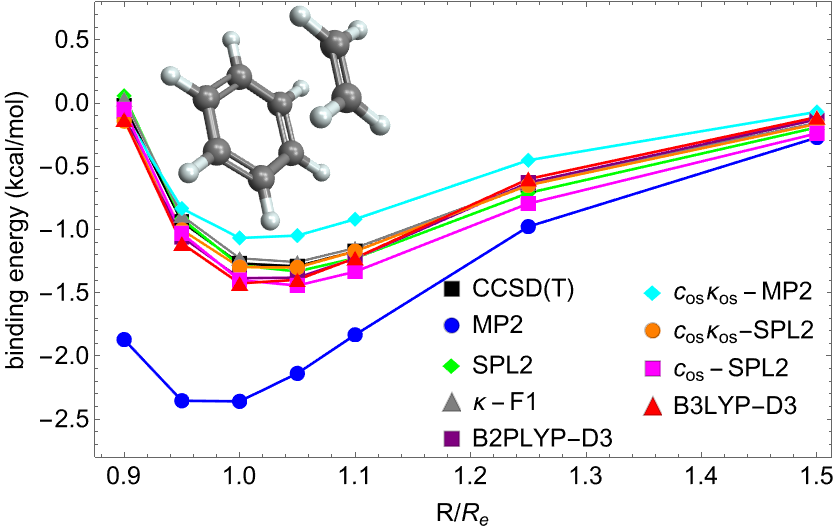}
\end{subfigure}
\begin{subfigure}{.49\textwidth}
    \centering
    \includegraphics[width=.95\linewidth]{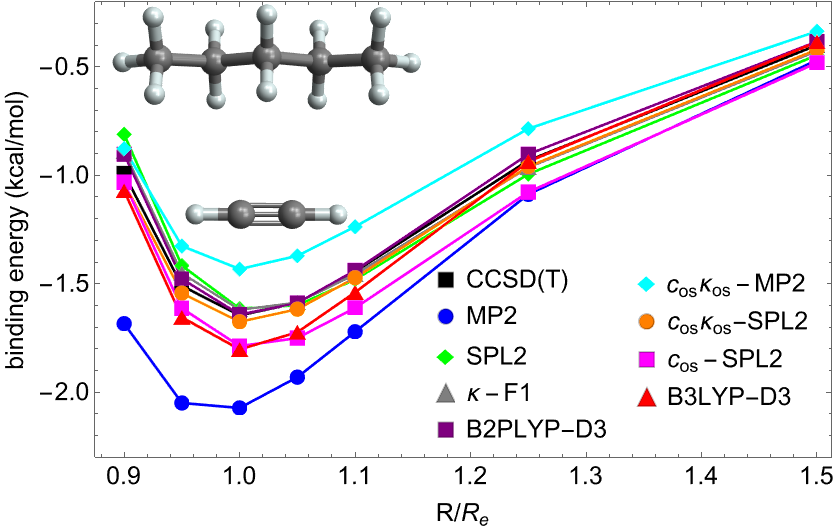}
\end{subfigure}
\caption{The dissociation curves of a dimer containing hydrogen bonds (top left), containing mixed interactions (top right) and two dimers containing dispersion interactions (bottom) for all 9 functionals and methods studied in the paper.}
\label{fig:dis_curves}
\end{figure*}

In Fig.~\ref{fig:Box_Pos}, we show the SPL2 correction for the S66x8 dissociation curves for small bond distances. We see that all that at $R=0.9R_e$, MP2 performs the worse but that SPL2 is able to reduce the median and reduce the amount of outliers. The same is true of the more accurate $c_{\rm os}\kappa_{\rm os}$-MP2 form which still has a quite large median but SPL2 is able to correct it as well. This shows that the SPL2 form is also able to correct the poor accuracy of MP2 for smaller bond lengths, although the methods are still more accurate at larger bond length such as the equilibrium distances or $R=1.05R_e$. However, the errors are not as bad as the original MP2 forms meaning that SPL2 can describe better these types of interactions at small bond distances.

\begin{figure*}
\begin{subfigure}{.49\textwidth}
\centering
\includegraphics[width=.95\linewidth]{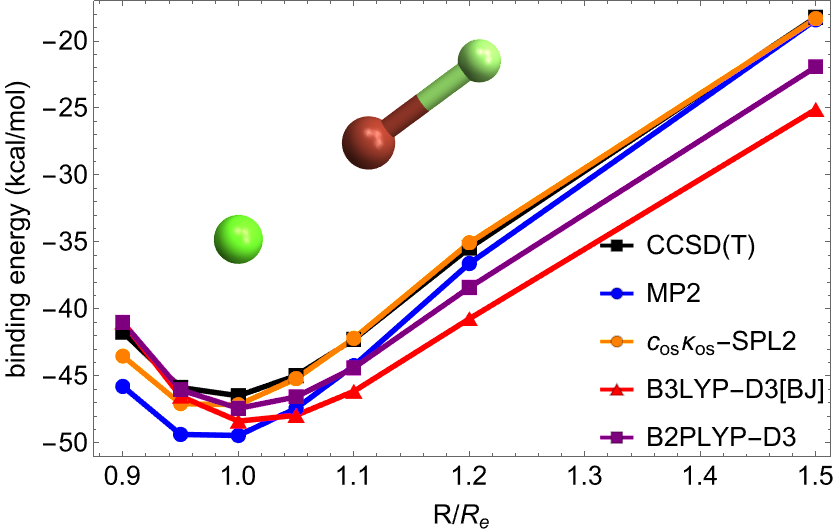}
\end{subfigure}
\begin{subfigure}{.49\textwidth}
\centering
\includegraphics[width=.95\linewidth]{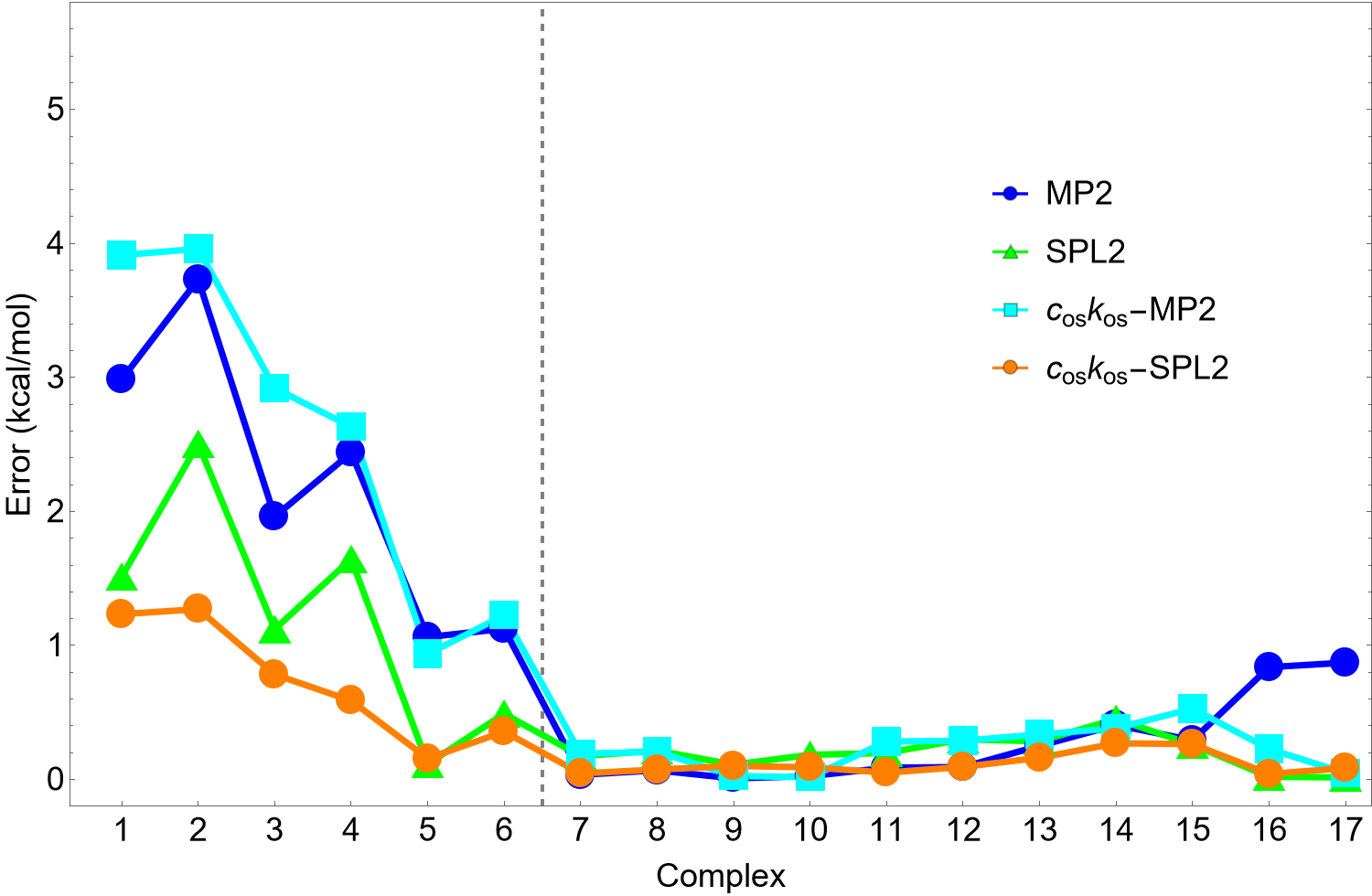}
\end{subfigure}
\caption{On the left the dissociation curve of \ce{NH3}-\ce{FBr} for CCSD(T), MP2, $c_{\rm os}\kappa_{\rm os}$-SPL2, B3LYP-D3[BJ] and B2PLYP-D3 and on the right the errors between the CCSD(T) reference data~\cite{OteJohDil-JCTC-14,RezRilHob-JCTC-12} and MP2, SPL2, $c_{\rm os},\kappa_{\rm os}$-MP2 and $c_{\rm os},\kappa_{\rm os}$-SPL2 respectively for the halogen bonded complexes of the B30~\cite{OteJohDil-JCTC-14,BauAlkFroElg-JCTC-13} and X40~\cite{RezRilHob-JCTC-12} datasets.}
\label{fig:b30dis2}
\end{figure*}
\bibliographystyle{achemso}
\bibliography{bib_clean}